\documentclass[twocolumn,pra,aps,showpacs,superscriptaddress,amssymb]{revtex4}
\usepackage{amsmath}
\usepackage{graphicx}
\newcommand{\tst}{\textstyle}

\newcommand{\mbf}{\mathbf}
\newcommand{\mrm}{\mathrm}
\newcommand{\ud}{\mathrm{d}}

\begin{document}

\title{Analytical solutions for the dynamics of two trapped interacting ultracold atoms}

\author{Zbigniew Idziaszek}
\affiliation{CNR-INFM BEC Center, I-38050 Povo (TN), Italy}
\affiliation{Centrum Fizyki Teoretycznej, Polska Akademia Nauk, 02-668 Warsaw, Poland}
\author{Tommaso Calarco}
\affiliation{CNR-INFM BEC Center, I-38050 Povo (TN), Italy}
\affiliation{ECT*, I-38050 Villazzano (TN), Italy}
\affiliation{ITAMP, Harvard Smithsonian Center for Astrophysics,
and Department of Physics, Harvard University, Cambridge, MA 02138, USA}

\begin{abstract}
We discuss exact solutions of the Schr\"odinger equation 
for the system of two ultracold atoms confined in an axially symmetric harmonic potential. 
We investigate different geometries of the trapping potential, in particular we study the 
properties of eigenenergies and eigenfunctions for quasi-one- and quasi-two-dimensional traps. 
We show that the quasi-one- and the 
quasi-two-dimensional regimes for two atoms can be already realized in the traps with moderately large 
(or small) ratios of the trapping frequencies in the axial and the transverse directions.
Finally, we apply our theory to Feshbach resonances for trapped atoms. Introducing in our description an 
energy-dependent scattering length we calculate analytically the eigenenergies for two trapped atoms in the 
presence of a Feshbach resonance.
\end{abstract}


\maketitle

\section{Introduction}
\label{Sec:Intro}

Atomic interactions 
represent one of the major ingredients for the schemes implementing quantum information processing 
in systems of neutral trapped atoms. Development of optical lattice technology \cite{Bloch}, 
atom chips \cite{microtraps} 
and dipole traps \cite{Grangier} allows to create tight external confinement for neutral atoms.
Feshbach resonances, widely used in recent experiments on ultracold atoms, permit for tuning of atomic
interactions, 
which was the key ingredient to accomplish molecular Bose-Einstein condensates and the superfluidity of
fermionic atoms in ultracold gases \cite{BEC-BCS}.
In addition, realization of Mott insulators \cite{Mott} gives the possibility to precisely control a number of 
atoms confined in a single well.
All these achievements make systems of ultracold neutral atoms very attractive in the context 
of quantum information processing or quantum control at the atomic level. 
Moreover, they also open a way for experimental studies of few-body interacting systems in the presence of tight 
external potentials.

On the theoretical level, the system of two interacting atoms in a harmonic trap can be solved analytically 
for spherically symmetric \cite{Busch} or axially symmetric harmonic potentials \cite{Idziaszek}. Both these
approaches model the interaction in terms of an $s$-wave delta pseudopotential \cite{Fermi,HuangPs}. Generalization
of the pseudopotential to higher partial waves \cite{Stock,IdziaszekPs} allows to solve the problem for 
generic types of (central) interactions, in the presence of   
spherically symmetric \cite{Stock} or axially symmetric \cite{IdziaszekPs} harmonic traps. Moreover, supplementing
the pseudopotential with an energy-dependent scattering length \cite{Blume,Bolda}
extends the validity of the analytic results to the case of very tight traps or large scattering lengths, and
accounts properly for the whole molecular spectrum \cite{Stock}. Such model provides for a very accurate 
description, which has been verified, for instance, in the recent experiment on the creation of molecules of fermionic
atoms in an optical lattice \cite{Stoferle}.

In this paper we discuss in detail the exact solutions for two interacting atoms confined in axially  
symmetric harmonic traps \cite{Idziaszek}. We present derivations of the analytical results discussing
different geometries of the trapping potential. In the limiting cases of the quasi-one- and quasi-two-dimensional
traps the system can be effectively described in terms of a lower-dimensional trap with renormalized scattering 
length. We investigate the limits of applicability of the quasi-one- and the quasi-two-dimensional descriptions
showing that they are valid already for moderately large (or small) ratios of the trapping frequencies in the axial 
and transverse directions. Finally, we consider the effects of Feshbach resonances on the trapped atoms.
Employing the standard theory of Feshbach resonances we express the energy-dependent scattering length in terms 
of the usual parameters describing the resonance. This allows to derive an explicit formula determining
the energy spectrum in the presence of a Feshbach resonance.

The paper is organized as follows. In section~\ref{Sec:SchrEq} we derive analytical solutions of 
the Schr\"odinger equation for 
two atoms confined in an axially symmetric trap, by expanding them in the basis of harmonic oscillator
wave functions. Section~\ref{Sec:Eigen} is devoted to the analysis of eigenenergies. In particular, in 
section~\ref{Sec:EnCig} we derive analytical results for cigar-shape traps, while the limiting case of 
the quasi-one-dimensional traps is analyzed in section~\ref{Sec:EnQ1D}. Section~\ref{Sec:EnPan} presents the 
analytical results for pancake-shape traps. The quasi-two-dimensional regime in these traps is studied in 
section~\ref{Sec:EnQ2D}. Section~\ref{Sec:Fun} analyzes the properties of eigenfunctions. In section~\ref{Sec:FunAn} 
we derive two series representations for the wave functions, which are valid for arbitrary ratio of radial to axial 
trapping frequencies. The behavior of the eigenfunctions in quasi-one- and quasi-two-dimensional traps
is discussed in sections~\ref{Sec:FunQ1D} and \ref{Sec:FunQ2D} respectively. In section~\ref{Sec:Feshbach} we 
illustrate the applicability of our theory, 
calculating the energy spectrum for two atoms interacting in the presence of a Feshbach resonance. 
We end in section~\ref{Sec:Conclusion} presenting some conclusions. Appendix~\ref{App:DetInt} 
presents some technical 
details related with the derivation of the energy spectrum for quasi-two-dimensional traps.

\section{System}
\label{Sec:SchrEq}

We consider two interacting atoms of mass $m$ confined in an axially symmetric harmonic trap with frequencies 
$\omega_{z}$ and $\omega_{\perp}$ in the axial and transverse directions, respectively.
The Hamiltonian of the system reads 
\begin{equation}
\label{Htot}
\hat{H} =  - \frac{\hbar^2}{2m} \nabla^2_1 - \frac{\hbar^2}{2m} \nabla^2_2 
+ V_\mrm{t}(\mathbf{r}_1) + V_\mrm{t}(\mathbf{r}_2) + V_\mrm{i}(\mathbf{r}_1 - \mathbf{r}_2),
\end{equation}
where $\mathbf{r}_1$ and $\mathbf{r}_2$ denote the positions of the two atoms, and $V_\mrm{t}(\mathbf{r})$ is the 
trapping potential
\begin{equation}
\label{Vtrap}
V_{t}(\mathbf{r}) = \frac{m}{2} ( \omega_\perp \rho^2 + \omega_z z^2 ),
\end{equation}
with $\rho^2 = x^2 + y^2$. For sufficiently low
energies, the scattering is purely of $s$-wave type and we model the interaction potential 
by a regularized delta function \cite{Fermi,HuangPs}
\begin{equation}
\label{Vi}
V_\mrm{i} (\mathbf{r}) = \frac{4 \pi \hbar^2 a}{m} \delta(\mathbf{r}) \frac{\partial}{\partial r} r,
\end{equation}
with $a$ denoting the $s$-wave scattering length.

For the harmonic trapping potential, the center-of-mass and relative motions are decoupled. 
Substituting new coordinates $\mbf{r}= \mbf{r}_1 - \mbf{r}_2$ and $\mbf{R}=(\mbf{r}_1 + \mbf{r}_2)/2$, we
decompose the total Hamiltonian into the center-of-mass part $\hat{H}_\mrm{CM}$ and relative
part $\hat{H}_\mrm{rel}$:
\begin{align}
\label{Hcm}
\hat{H}_\mrm{CM} = & - \frac{\hbar^2}{2M} \nabla^2_{R} + \frac{M}{m} 
V_{t}(\mathbf{R}) \\
\label{Hrel}
\hat{H}_\mrm{rel} = & - \frac{\hbar^2}{2\mu} \nabla^2_{r} + \frac{\mu}{m}
V_{t}(\mathbf{r}) + V_\mrm{i} (\mathbf{r}),
\end{align}
where $\mu =m/2$ and $M =2m$ denote the reduced and the total mass, respectively.  

In the following we use dimensionless variables, in which all
lengths are expressed in units of 
$d=\sqrt{\hbar/(\mu \omega_z)}$,
and all energies are expressed in units of $\hbar \omega_z$.
The eigenfunctions of the center-of-mass Hamiltonian $\hat{H}_\mrm{CM}$
are the usual harmonic-oscillator wave functions. The eigenfunctions
$\Psi(\mbf{r})$ of the relative motion have to be determined from 
\begin{equation}
\label{SchrEq}
\left[- \frac{1}{2} \nabla^2_{r} + \frac12 \left( \eta^2 \rho^2 + z^2\right)+
2 \pi a \delta(\mathbf{r}) \frac{\partial}{\partial r} r
\right]\Psi(\mbf{r}) = E \Psi(\mbf{r})
\end{equation}
where $\eta = \omega_{\perp}/\omega_z$.
Its solutions can be found by decomposing $\Psi(\mbf{r})$ into the complete set 
of the harmonic oscillator wave functions \cite{Busch}
\begin{equation}
\label{PsiDecomp}
\Psi(\mbf{r}) = \sum_{n,k} c_{n,k}  \Phi_{n,0}(\rho,\varphi) \Theta_{k}(z).
\end{equation}
where, $\Phi_{n,m}(\rho,\varphi)$ denotes the states of the two-dimensional harmonic oscillator
in polar coordinates $(\rho,\varphi)$ with the radial and angular quantum numbers $n$ and $m$, 
respectively, whereas $\Theta_{k}(z)$ is the one-dimensional harmonic-oscillator wave function with the 
quantum number $k$. 
We note, that only the states with $m=0$, i.e. the states with vanishing angular momentum along $z$ 
enter the summation in Eq. (\ref{PsiDecomp}). The states with
$m \neq 0$ vanish at $\mbf{r} = 0$, and they are not perturbed in the presence of the interaction potential. 
Substituting the expansion (\ref{PsiDecomp}) into the Schr\"odinger equation (\ref{SchrEq}) yields 
\begin{align}
\label{SchrEq1}
0 = &  \sum_{n,k} c_{n,k} (E_{n,k} - E) \Phi_{n,0}(\rho,\varphi) \Theta_{k}(z) & \\
& + 2 \pi a \delta(\mathbf{r}) \frac{\partial}{\partial r} r \sum_{n,k} c_{n,k} 
\Phi_{n,0}(\rho,\varphi) \Theta_{k}(z),
\end{align}
where $E_{n,k}=1/2+ \eta + k + 2 \eta n$ are dimensionless eigenenergies of the
three-dimensional axially-symmetric harmonic oscillator.
To determine the expansion coefficients $c_{n,k}$ we project Eq. (\ref{SchrEq1}) onto state 
$\Phi_{n^{\prime},0}(\rho,\varphi) \Theta_{k^{\prime}}(z)$ with arbitrary
$n^{\prime}$ and $k^{\prime}$, obtaining 
\begin{equation}
\label{cnk}
c_{n,k} = {\cal C} \frac{\Phi_{n,0}^{\ast}(0,\varphi) \Theta_{k}^{\ast}(0)}{E_{n,k} - E},
\end{equation}
where ${\cal C}$ is a constant fixed by the normalization of the wave function. The value of
${\cal C}$ is related to the expansion coefficients $c_{n,k}$ through 
\begin{equation}
\label{C}
{\cal C} = 2 \pi a \left[ \frac{\partial}{\partial r} \left( r \sum_{n,k} c_{n,k}
\Phi_{n,0}(\rho,\varphi) \Theta_{k}(z) \right) \right]_{r=0}
\end{equation}
Substituting the solution (\ref{cnk}) for coefficients $c_{n,k}$ into Eq. (\ref{C}), the numerical constant
${\cal C}$ disappears, and we obtain an equation which determines the eigenenergies with $m=0$:
\begin{equation}
\label{Energ}
- \frac{1}{2 \pi a} = \left[ \frac{\partial}{\partial r} r \Psi_{\cal E}(\mbf{r})
\right]_{r=0},
\end{equation}
where
\begin{equation}
\label{DefPsiE}
\Psi_{\cal E}(\mbf{r}) \equiv \sum_{n,k} 
\frac{\Phi_{n,0}^{\ast}(0,\varphi) \Phi_{n,0}(\rho,\varphi)
\Theta_{k}^{\ast}(0) \Theta_{k}(z)}{2 \eta n + k - {\cal E}},
\end{equation}
and ${\cal E} = E -E_0$ denotes the energy shifted by the zero-point oscillation energy
$E_0 = 1/2 + \eta$. For values of ${\cal E}$ solving Eq. (\ref{Energ}), 
the functions $\Psi_{\cal E}(\mbf{r})$ represent the non-normalized eigenstates of the relative motion. 
We note that $\Psi_{\cal E}(\mbf{r})$ is proportional to the single-particle Green 
function $G(0,\mbf{r})$ of the three-dimensional anisotropic harmonic oscillator.
We stress that, the regularization operator $\frac{\partial}{\partial r} r$ and the summation 
in Eqs. (\ref{Energ}) and (\ref{DefPsiE}) cannot be interchanged, 
and the summation over $n$ and $k$ must be done first. This is related with the divergent behavior of the Green function at small $r$, which is regularized by $\frac{\partial}{\partial r} r$.
To perform the summation in Eq. (\ref{DefPsiE}) we express 
the denominator in Eq. (\ref{DefPsiE}) in terms of the following integral:
\begin{equation}
\label{Int} 
\frac{1}{2 \eta n + k -{\cal E}} = \int_0^{\infty} \ud t \ e^{-t (2 \eta n + k -{\cal E})}.
\end{equation}
Since $n \geq 0$ and $k \geq 0$, the integral representation (\ref{Int}) is valid for ${\cal E}<0$. The wave function of the two-dimensional harmonic-oscillator with $m=0$ is given by  
\begin{equation}
\Phi_{n,0}(\rho,\varphi) = \frac{\sqrt{\eta}}{\sqrt{\pi}} e^{-\eta \rho^2/2} L_n(\eta \rho^2)
\end{equation}
where $L_n(x)$ is the Laguerre polynomial. To perform the summation over $n$ in Eq. (\ref{DefPsiE}), we 
utilize the fact that $L_n(0)=1$ and we apply the generating function for Laguerre polynomials \cite{Gradshteyn}
\begin{equation}
\label{SumL}
\sum_{n=0}^{\infty} L_n(x) z^n = (1-z)^{-1} \exp \left(\frac{xz}{z-1}\right)
\end{equation} 
On the other hand, the sum over $k$ involves the eigenstates of the one-dimensional harmonic-oscillator 
\begin{equation}
\Theta_{k}(z) = \frac{e^{-z^2/2}}{\pi^{1/4} \sqrt{2^k k!}} H_k(z),
\end{equation}
where $H_k(z)$ is the Hermite polynomial. The summation can be performed with the help of the following generating 
function for the products of Hermite polynomials \cite{Prudnikov}
\begin{equation}
\label{SumH}
\sum_{k=0}^{\infty} \frac{t^k}{2^k k!} H_k(x) H_k(y) = \frac{e^{(2txy-t^2 x^2- t^2 y^2)/(1-t^2)}}{\sqrt{1-t^2}} 
\end{equation}
Upon inserting Eq. (\ref{Int}) into (\ref{DefPsiE}), and performing the summation according to 
(\ref{SumL}) and (\ref{SumH}), we find the following integral representation for $\Psi_{\cal E}(\mbf{r})$
\begin{align}
\label{PsiE} 
\Psi_{\cal E}(\mbf{r}) = & \frac{\eta}{(2 \pi)^{\frac{3}{2}}}
\int_{0}^{\infty} \! \! \! \! \ud t \ \frac{\exp\!\left[ t E
-\frac{z^2}{2} \coth t - \frac{\eta \rho^2}{2} \coth(\eta t)
\right]} {\sqrt{\sinh(t)} \sinh (\eta t)}.
\end{align}
The integral is convergent for $E<E_0$, however, the validity of the final result will be extended to 
energies $E > E_0$, by virtue of the analytic continuation.

Let us investigate now the behavior of $\Psi_{\cal E}(\mbf{r})$ for small values of $\mbf{r}$. In the limit 
$\mbf{r} \rightarrow 0$, the main contribution to the integral comes from small arguments $t$. 
In the leading order we can neglect the dependence on the energy $E$, and the expansion of (\ref{PsiE})
for small $t$ yields
\begin{equation}
\label{PsiAppr}
\Psi_{\cal E}(\mbf{r}) \approx \frac{1}{(2 \pi)^{3/2}} \int_{0}^{\infty} \! \! \! \! \ud t \ 
\frac{e^{-r^2/(2t)}}{t^{3/2}} = \frac{1}{2 \pi r}, \qquad r \ll 1.
\end{equation}
We note, that, for small $r$ the function $\Psi_{\cal E}(\mbf{r})$ diverges in the same way as the Green function in 
a homogeneous space. This is related with the fact that at short distances the behavior of the wave function is
determined mainly by the interaction between the particles. After extracting the divergent
behavior of $\Psi_{\cal E}(\mbf{r})$, we can simplify Eq. (\ref{Energ}) determining the eigenenergies. 
To this end, we substitute the integral representation (\ref{PsiE}) into (\ref{Energ}), 
and subtract from $\Psi_{\cal E}(\mbf{r})$ the r.h.s. of (\ref{PsiAppr}). This can be done since the term $1/(2 \pi r)$
is removed by the regularization operator, and does not give any contribution to (\ref{Energ}).
In this way we obtain a simpler expression determining the energy levels for $m=0$
\begin{equation}
\label{Energ1}
- \frac{\sqrt{\pi}}{a} = {\cal F}(- {\cal E}/2),
\end{equation}
where
\begin{equation}
\label{Deff}
{\cal F}(x) = 
\int_{0}^{\infty} \! \! \! \! \ud t \! \left[ \frac{ \eta e^{- x t}}
{\sqrt{1-e^{-t}} \left(1 - e^{-\eta t}\right)} - \frac{1}{t^{3/2}} \right] \quad \textrm{for } x>0 
\end{equation}
Similarly to \eqref{PsiE}, the validity of the integral representation \eqref{Deff} is limited to ${\cal E} < 0$. In
general, ${\cal F}(x)$ can be calculated from the following series representation, valid for all values of $x$:
\begin{equation}
\label{FSer}
{\cal F}(x) = \frac{\eta}{2 \pi} \sum_{n=0}^{\infty} \left( \frac{\Gamma( x + n \eta)}{\Gamma(\frac12 +x + n \eta)}
- \frac{1}{\sqrt{\eta}\sqrt{n+1}} \right) + \frac{\sqrt{\eta}}{2 \pi} \zeta({\tst \frac12}),
\end{equation}
where $\Gamma(x)$ is the Gamma function and $\zeta({\tst \frac12})$ denotes the Riemann zeta function.
To derive \eqref{FSer} we retain the summation over $n$ in Eq.~\eqref{DefPsiE}, while we perform the 
integration over $t$, and the rest of the steps is the same as in derivation of \eqref{Deff}.

Finally, in the numerical calculations of ${\cal F}(x)$ we were using a recurrence relation,
which relates the values of function ${\cal F}(x)$ for different arguments $x$. To derive this formula 
we calculate the difference between ${\cal F}(x)$ and ${\cal F}(x+\eta)$ obtaining with the 
help of \eqref{Deff}
\begin{align}
{\cal F}(x)-{\cal F}(x+\eta) = \int_{0}^{\infty} \!\!\!\! \ud t \ \frac{\eta e^{- x t}}{\sqrt{1-e^{-t}}} 
= \eta \sqrt{\pi} \frac{\Gamma(x)}{\Gamma(x+\frac{1}{2})} 
\label{fDiff1}
\end{align}

\section{Eigenenergies}
\label{Sec:Eigen}

\subsection{Cigar shape traps}
\label{Sec:EnCig}

\begin{figure}
	 \includegraphics[width=8cm,clip]{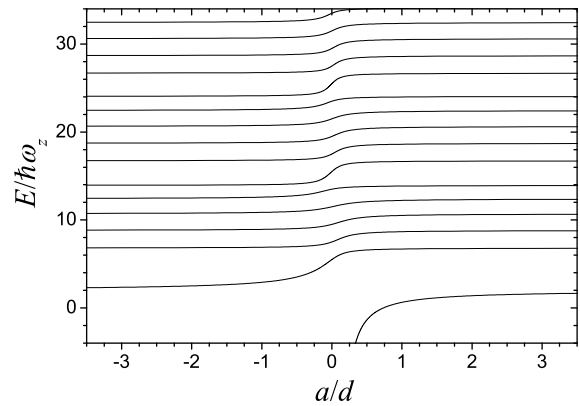}
	 \caption{
	 \label{fig:E5}
	 Eigenenergies of the relative motion for two atoms interacting via $s$-wave pseudopotential and confined
	 in a harmonic trap with $\eta = \omega_{\perp}/\omega_z = 5$. 
	 The scattering length $a$ is scaled in units of the harmonic  
	 oscillator length $d = \sqrt{\hbar / (\mu \omega_z)}$. 
	 }
\end{figure}

In this section we discuss the energy spectrum of two interacting atoms confined in a trap with $\eta>1$. 
Let us first consider the case $\eta=n$, with $n$ being a positive integer. In this particular case we can treat the
function $1-e^{-n t}$, which appears in the denominator of the integrand in Eq. (\ref{Deff}), as a polynomial of 
order $n$ in the variable $e^{-t}$, and perform a decomposition into a sum of simple fractions 
\begin{equation}
\frac{1}{1-e^{- n t}} = \frac{1}{n} \sum_{m=0}^{n-1} \frac{1}{1-e^{-t - i 2\pi m/n}}
\end{equation}
Substituting this decomposition into (\ref{Deff}) yields 
\begin{align}
 {\cal F}(x) = & \int_{0}^{\infty} \!\!\!\! \ud t \ \left[ \frac{ e^{- x t}}
{\left(1 - e^{-t}\right)^{3/2}} - \frac{1}{t^{3/2}} \right] \nonumber \\
& {} + \sum_{m=1}^{n-1} \int_{0}^{\infty} \ud t \ \frac{ e^{- x t}}
{\sqrt{1-e^{-t}} \left(1-e^{-t - i 2\pi m/n}\right)} \nonumber \\
= & - 2 \sqrt{\pi} \frac{\Gamma(x)}{\Gamma(x-\frac{1}{2})} \nonumber \\
\label{fcEx}
& {} + \sqrt{\pi} \frac{\Gamma(x)}{\Gamma(x+\frac{1}{2})}
\sum_{m=1}^{n-1} F\left(1,x;x+{\tst \frac{1}{2}};e^{i \frac{2\pi m}{n}}\right),
\end{align}
where $F(a,b;c;x)$ denotes the hypergeometric function. We note that Eq. (\ref{fcEx}) is derived from
the integral representation (\ref{Deff}) applicable for $x>0$, however the validity of the final result can be extended
to $x \leq 0$ by virtue of the analytic continuation. Despite the presence of the complex roots of $1$
in the argument of the hypergeometric function, it can be easily verified that 
the whole expression remains real for $x \in \mathbf{R}$.
For the special case $n=1$, the second term in Eq. (\ref{fcEx}) disappears and we obtain the well-known result for
the spherically symmetric trap \cite{Busch,Block}
\begin{equation}
\label{fSph}
{\cal F}(x) = - 2 \sqrt{\pi} \frac{\Gamma(x)}{\Gamma(x-\frac{1}{2})}, \qquad \eta=1.
\end{equation}
Fig.~\ref{fig:E5} presents the energy levels calculated for $\eta=5$ from Eq. (\ref{Energ1}), 
with ${\cal F}(x)$ given by the exact formula (\ref{fcEx}). For $a=0$ the eigenvalues are given by the poles of 
${\cal F}(x)$, and we recover obviously the energy spectrum of the harmonic oscillator. On the other hand,
the eigenvalues for $a \rightarrow + \infty$ and for 
$a \rightarrow - \infty$ approach the same asymptotic values, corresponding to zeros of ${\cal F}(x)$. 
The level spacing for large $a$ is not uniform, as in the case of a spherically symmetric trap \cite{Busch}, 
and the distance 
between energy levels is larger every fifth level, which results from the geometry of the trap. For $a>0$
(repulsive potential) the energy levels are shifted upward with respect to the non-interacting case, while for $a<0$ (attractive potential) they are shifted downwards. In a homogeneous space, the three-dimensional regularized delta potential possesses a single bound state for $a>0$, with energy $E=-\hbar^2 /m a^2$. From 
Fig.~\ref{fig:E5} we see that such a state is also present in the case of harmonic confinement, however, its energy 
is shifted upward due to the presence of the trap. Moreover, the presence of external confinement gives rise to the 
appearance of a bound-state also for $a<0$, which can be observed in Fig.~\ref{fig:E5} as a branch of the spectrum 
starting from the energy of zero-point oscillations.  

\begin{figure}
	 \includegraphics[width=8cm,clip]{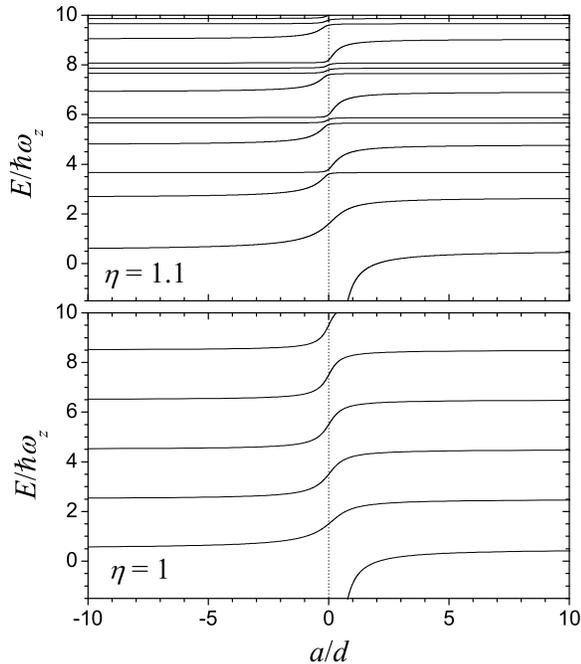}
	 \caption{
	 \label{fig:E1.1}
	 Eigenenergies of the relative motion for two atoms interacting via $s$-wave pseudopotential and confined
	 in a harmonic trap with $\eta = \omega_{\perp}/\omega_z = 1.1$ (upper plot) and 
	 $\eta = 1$ (lower plot). The scattering length $a$ is scaled in units of the harmonic  
	 oscillator length $d = \sqrt{\hbar / (\mu \omega_z)}$.
	 }
\end{figure}

Fig.~\ref{fig:E1.1} presents the exact energy levels for $\eta=1.0$ (upper plot) and $\eta=1.1$ (lower plot). 
The former are given by the analytical result (\ref{fSph}), while the latter were calculated numerically from 
Eq.~(\ref{Deff}) for ${\cal E}<0$, and with the help of the recurrence relation (\ref{fDiff1}) for ${\cal E} \geq 0$. 
Comparing upper and
lower plots, we note that the energy spectrum for $\eta=1.1$ has a richer structure 
than the one for the spherically 
symmetric trap ($\eta=1$). In the latter case some of the excited states are degenerate, and they do not appear
in the energy spectrum given by Eq.~\eqref{Energ1}. As can be easily verified by taking linear combinations 
of degenerate wave functions, the number of solutions not vanishing at $\mbf{r}=0$ can be always reduced to one, 
and as a consequence only one of the degenerate states is affected by zero-range interaction. 
Analyzing the behavior of the eigenenergies close to $a=0$, 
we notice the appearance of avoided crossings for non-integer values of $\eta$.

\subsection{Quasi-1D regime}
\label{Sec:EnQ1D}

Now, we turn to the discussion of the energy spectrum for $\eta \gg 1$. For $x \sim \eta$
and $\eta \gg 1$, the main contribution to the integral in (\ref{Deff}) comes from small 
arguments $t$. In this regime we perform the approximation $\sqrt{1-e^{-t}} \approx \sqrt{t}$ 
in the denominator of (\ref{Deff}) and obtain
\begin{equation}
\label{fapprInt}
{\cal F}(x) \approx 
\int_{0}^{\infty}  \! \! \! \!  \ud t \ \left[ \frac{ \eta e^{- x t}}
{\sqrt{t} \left(1 - e^{-\eta t}\right)} - \frac{1}{t^{3/2}} \right], \quad \eta \gg 1
\end{equation}
Now the integration can be done analytically. This yields 
\begin{equation}
\label{fappr}
{\cal F}(x) \approx 
\sqrt{\pi \eta} \, \zeta_\mrm{H}\left(1/2,x/\eta\right), \quad x \gtrsim \eta 
\end{equation}
where $\zeta_\mrm{H}(s,a)$ denotes the Hurwitz Zeta function: 
$\zeta_\mrm{H}(s,a) =  \sum_{k=0}^{\infty} (k+a)^{-s}$ 
\cite{Elizalde}.
To extend the validity of the approximate result (\ref{fappr}) to $x$ positive and much smaller
than $\eta$, or to negative $x$, 
we make use of the recurrence relation \eqref{fDiff1}. Applying the recurrence formula once, we find 
\begin{equation}
\label{fappr1}
{\cal F}(x) \approx \sqrt{\pi \eta} \, \zeta_H\left(\frac{1}{2},1+\frac{x}{\eta}\right) 
+ \eta \sqrt{\pi} \frac{\Gamma(x)}{\Gamma(x+\frac{1}{2})}.
\end{equation}
Numerical comparison of the exact result (\ref{fcEx}) with the approximation (\ref{fappr1}) shows 
that the latter provides quite accurate values for $x > -\eta$. Thus, utilizing (\ref{fappr1})  
we are able to calculate the energy of the ground state and of the first excited states, up to ${\cal E}<2 \eta$. 
To find the energies of higher excited states one can apply recursively (\ref{fDiff1}).

It is interesting to compare our results with the predictions of one-dimensional model, where the 
scattering length is renormalized due to the tight confinement in the transverse direction
\cite{Olshanii}. At low energies of the scattered particles, 
the one-dimensional pseudopotential takes the form \cite{Olshanii}
\begin{equation}
\label{V1D}
V_\mrm{1D}(r) = - \frac{\hbar^2}{\mu a_\mrm{1D}} \delta(r),
\end{equation}
where $a_\mrm{1D}$ is the one-dimensional scattering length. Repeating similar steps as for the three-dimensional 
trap, we obtain the following implicit equation determining the eigenenergies 
of the relative motion in a one-dimensional harmonic potential \cite{Busch}
\begin{equation}
\label{En1D}
2 a_\mrm{1D} = \frac{\Gamma(-\frac{{\cal E}}{2})}{\Gamma(-\frac{{\cal E}}{2}+\frac{1}{2})}
\end{equation}
We compare the latter equation with the three-dimensional result for $\eta \gg 1$,
\begin{equation}
\label{EnQ1D}
- \frac{1}{a} = \sqrt{\eta} \, \zeta_\mrm{H}\left(\frac{1}{2},1-\frac{{\cal E}}{2 \eta}\right) 
+ \eta \frac{\Gamma(-\frac{{\cal E}}{2})}{\Gamma(-\frac{{\cal E}}{2}+\frac{1}{2})},
\end{equation}
obtained by substituting into Eq. (\ref{Energ1}) the function ${\cal F}(x)$ given by (\ref{fappr1}). 
For the energies $|{\cal E}| \ll \eta$ one can neglect 
the dependence on energy in the first term on the r.h.s. of (\ref{EnQ1D}):
$\zeta_\mrm{H}\left(1/2,1-{\cal E}/(2 \eta)\right) \approx \zeta\left(1/2\right)$.
Then it is straightforward to observe that Eqs. (\ref{En1D}) and (\ref{EnQ1D}) give the same energy spectrum, provided
the one-dimensional scattering length $a_\mrm{1D}$ is related to $a$ by 
\begin{equation}
\label{a1D}
a_\mrm{1D} = -\frac{1}{2 \eta a} - \frac{\zeta(\frac{1}{2})}{2 \sqrt{\eta}}.
\end{equation}
Expressing this relation in physical units, we recover the result of Ref. \cite{Olshanii} 
\begin{equation}
\label{a1Dph}
a_\mrm{1D} = -\frac{d_{\perp}^2}{2 a} - \frac{d_{\perp}}{2} \zeta({\tst \frac{1}{2}}),
\end{equation}
where $d_{\perp} = \sqrt{\hbar /(\mu \omega_{\perp})}$. 

In the previous works \cite{BoldaQ,Idziaszek}
it was shown that a one-dimensional model with renormalized scattering length provides a very 
accurate description of the spectrum for $E>E_0$. On the other hand, this approach is not valid 
for energies $E<E_0$ when the system possesses a bound state. Its energy has to be determined
from Eqs. (\ref{Energ1}) and (\ref{fappr}), derived from the three-dimensional approach.
The condition for the energy of a bound state expressed in the physical units reads 
\begin{equation}
\label{En1DB}
-\frac{d_{\perp}}{a} = \zeta_\mrm{H}\left(\frac{1}{2},\frac{E_0-E}{2 \hbar \omega_{\perp}}\right),
\end{equation}
which agrees with the results of \cite{Bergeman}. We note that Eq.~\eqref{En1DB} involves only the 
trapping frequency $\omega_{\perp}$ in the tightly confined direction. Due to this reason 
the one-dimensional model, which depends crucially on $\omega_{z}$, fails to 
describe the energy of a bound state.

\begin{figure}
	 \includegraphics[width=8.5cm,clip]{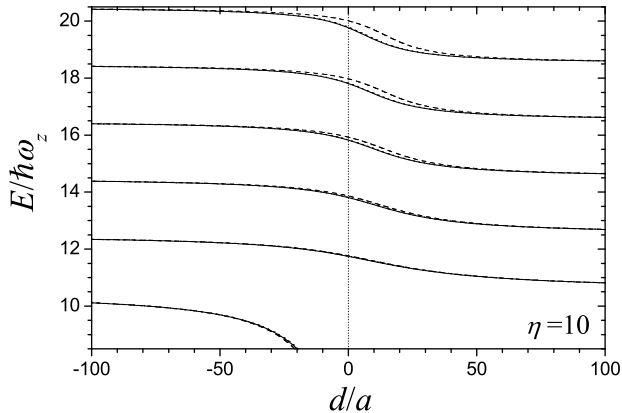}
	 \caption{
	 \label{Fig:E10}
	 Eigenenergies of the relative motion for two atoms interacting via $s$-wave pseudopotential and confined
	 in a harmonic trap with $\eta = \omega_{\perp}/\omega_z = 10$.
	 The exact energy levels (solid lines) are compared with 
	 predictions for the one-dimensional model with energy-dependent 
	 (dotted line, indistinguishable from the solid one) and with the standard, energy-independent
	 renormalized scattering length (dashed lines).
	 The three-dimensional scattering length $a$ is scaled by the 
	 harmonic oscillator length $d = \sqrt{\hbar / (\mu \omega_z)}$.  
	 }
\end{figure}

We have verified that the quasi-one-dimensional regime for two interacting atoms does not require $\eta$ very large, but is already realized for $\eta \sim 10$.
The case of $\eta = 10$ is illustrated in Fig.~\ref{Fig:E10}, where we compare the exact energy
levels with the ones calculated from Eq.~\eqref{En1D} for the one-dimensional spectrum with $a_\mrm{1D}$ given by 
\eqref{a1D}. We observe quite good agreement for the lowest eigenstates. For higher excited states the two 
approaches start to differ around the unitarity point ($1/a=0$). The predictions of the one-dimensional model can be improved when the eigenenergies are calculated assuming an 
energy-dependent one-dimensional scattering length \cite{Bolda}
\begin{equation}
\label{a1DE}
a_\mrm{1D}(E) = -\frac{d_{\perp}^2}{2 a} - \frac{d_{\perp}}{2} \zeta_\mrm{H}\left(\frac{1}{2},
\frac{E_0-E}{2\hbar \omega_\perp}\right),
\end{equation}
This expression can be obtained from the result \eqref{fappr}; however, in this case we do not apply 
the approximation ${\cal E}=0$ for the first term of \eqref{fappr}. We note an excellent agreement between 
the exact energy
levels and the one-dimensional spectrum evaluated with $a_\mrm{1D}(E)$, which on the scale of Fig.~\ref{Fig:E10}
are indistinguishable.

\subsection{Pancake shape traps}
\label{Sec:EnPan}

In this section we investigate the energy spectrum of the two interacting atoms 
confined in harmonic traps with $0<\eta<1$. First, we derive an explicit formula for 
${\cal F}(x)$ in the case when $\eta$ is the inverse of a positive integer.
We start from the integral representation (\ref{Deff}) of
function ${\cal F}(x)$, which for $\eta=1/n$ takes the following form:
\begin{equation}
\label{f3}
{\cal F}(x) = 
\int_{0}^{\infty} \!\!\!\! \ud t \ \left[ \frac{ e^{- x t}}
{n \sqrt{1-e^{-t}} \left(1 - e^{-t/n}\right)} - \frac{1}{t^{3/2}} \right]. 
\end{equation}
To calculate the integral we make use of the following identity 
\begin{equation}
\label{Ident}
\frac{1}{1-e^{-t/n}} = \frac{1}{1-e^{-t}} \sum_{m=0}^{n-1} e^{-t m /n}.
\end{equation}
To prove this identity one can use the formula $1-x^n=(1-x)(1+x+x^2+\ldots+x^{n-1})$ with $x=e^{-t/n}$.
Substituting (\ref{Ident}) into (\ref{f3}) leads to 
\begin{equation}
\label{f4}
{\cal F}(x) = \frac{1}{n} \sum_{m=0}^{n-1}
\int_{0}^{\infty} \!\!\!\! \ud t \ \left[ \frac{ e^{- t (x+m/n) }}
{\left(1-e^{-t}\right)^{3/2}} - \frac{1}{t^{3/2}} \right], 
\end{equation}
which can be evaluated analytically
\begin{equation}
\label{fpEx}
{\cal F}(x) = - \frac{2 \sqrt{\pi}}{n} \sum_{m=0}^{n-1}
\frac{\Gamma(x+\frac{m}{n})}{\Gamma(x-\frac{1}{2}+\frac{m}{n})}.
\end{equation}
We note that for the special case of a spherically symmetric trap ($n=1$), 
we recover obviously the result (\ref{fSph}).

\begin{figure}
	 \includegraphics[width=8cm,clip]{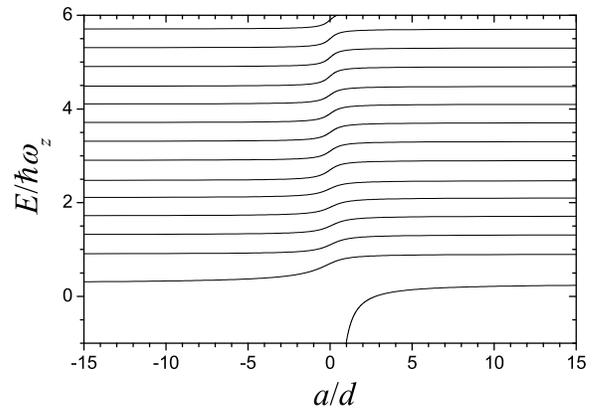}
	 \caption{
	 \label{fig:E02}
	 Eigenenergies of the relative motion for two atoms interacting via $s$-wave pseudopotential and confined
	 in a harmonic trap with $\eta = \omega_{\perp}/\omega_z = 1/5$. 
	 The scattering length $a$ is scaled in units of the harmonic  
	 oscillator length $d = \sqrt{\hbar / (\mu \omega_z)}$. 
	 }
\end{figure}

The exact energy spectrum calculated from combined Eqs.~(\ref{Energ1}) and (\ref{fpEx}), 
and for $\eta=1/5$ is presented in Fig.~\ref{fig:E02}. 
We observe that the energy spectrum has similar features as for the cigar shape traps.
In the limit of $a \rightarrow \pm \infty$ the eigenenergies approach the same asymptotic values,
irrespective of the sign of the scattering length. In contrast to the cigar shape traps, the level spacing at 
$a \rightarrow \pm \infty$ is almost uniform, except for the gap between the ground and the first-excited state.
Again, due to the external confinement, the solutions with energy $E<E_0$, corresponding to the 
bound states of the interaction potential,
occur both for positive and negative values of the scattering length.

\subsection{Quasi-2D regime}
\label{Sec:EnQ2D}

Now, we turn to the analysis of the energy spectrum in the quasi-two-dimensional traps with $\eta \ll 1$.
For simplicity we assume $\eta=1/n$ with $n$ being an integer, however, this choice is not crucial for the 
applicability of the final results, as we show later. We start our derivation from 
formula (\ref{fpEx}), in which for $n \gg 1$ we approximate the summation by an integration 
\begin{equation}
\label{fpAp1}
{\cal F}(x) \approx \int_{0}^{1} \! \! \! \ud t \ B\left(x+t,{\tst -\frac{1}{2}}\right), \quad \eta \ll 1,
\end{equation}
where $B(x,y) = \Gamma(x) \Gamma(y) /\Gamma(x+y)$ is the Euler beta function. This approximation is valid for $x>0$,
which guarantees that the function $B\left(x+t,-\frac{1}{2}\right)$ is free from the singularities
in the interval of integration. As we show in Appendix~\ref{App:DetInt}, the integral in Eq. (\ref{DefPhi}) 
can be expressed in terms of the following functions:
\begin{equation}
\label{DefPhi}
\int_{0}^{1} \! \! \! \ud t \ B\left(x+t,{\tst -\frac{1}{2}}\right) = - \Phi (x) - \ln x,
\end{equation}
with 
\begin{align}
\label{PhiSer}
\Phi (x) = & 2 - \log (1+x) \nonumber \\
& {} + 2 \sum_{k=1}^{\infty} \frac{(2k)!}{(2^k k!)^2} \left[(k+{\tst \frac{1}{2}})\log\frac{x+k}{x+k+1}+1\right].
\end{align}
Eq. (\ref{Energ1}) together with the approximate result (\ref{fpAp1}) determine the energy of the bound state 
(${\cal E} < 0$) in the quasi-two-dimensional traps. Expressing this result in the physical units we obtain
\begin{equation}
 \frac{\sqrt{\pi} d}{a} = \Phi \left(\frac{E_0-E}{2 \hbar \omega_z} \right)+ \ln 
\left( \frac{E_0-E}{2 \hbar \omega_z} \right).
\end{equation}
We observe that in quasi-two-dimensional traps the properties of a bound-state 
depend solely on the trap frequency in the tightly confined axial direction. For a shallow bound state
($E_0-E \ll \hbar \omega_z$) we can approximate $\Phi (-{\cal E}/2)$ by $\Phi(0)\approx 1.938$, and
in this regime we recover the result of Ref.~\cite{Petrov2}: $E_0 - E =\hbar \omega_z 0.288 \exp( \sqrt{\pi}d/a)$.

Let us investigate now the energy spectrum for ${\cal E} > 0$ containing the excited states. In the following, 
we will compare our results, obtained from the three-dimensional description with predictions 
for the two-dimensional system, with the scattering length renormalized due to the tight confinement 
in the axial direction \cite{Petrov1,Petrov2}.
In analogy to the derivation of Huang and Yang \cite{HuangPs}, one can show 
that in two dimensions the $s$-wave pseudopotential takes the form 
\begin{equation}
\label{V2D}
V_\mrm{2D}(r)= - \frac{\pi \hbar^2}{\mu \ln (k a_\mrm{2D})} \delta(\mbf{r})\left(1 - \ln({\cal A} k \rho) \rho 
\frac{\partial}{\partial \rho}\right),
\end{equation}
where $k^2=2 \mu E/\hbar^2$, ${\cal A} = e^\gamma/2$, with $\gamma$ denoting the Euler constant
and $a_{2D}$ is a two-dimensional scattering length related with the 
the {\it s}-wave scattering phase shift $\delta_0$ by $\tan \delta_0 = (\pi/2) \ln^{-1} (k a_{2D})$.
The regularization operator $(1 - \ln({\cal A} k \rho) \rho \frac{\partial}{\partial \rho})$ 
removes the logarithmic-type divergence of the two-dimensional scattering solution at $\rho=0$ 
\cite{Wodkiewicz}. We note that even in the limit $k \rightarrow 0$,
the pseudopotential (\ref{V2D}) depends on energy.

Using similar techniques as for the three-dimensional trap, 
one can show that the energy spectrum of two interacting atoms 
confined in the two-dimensional harmonic trap of frequency $\omega_{\perp}$
is given by \cite{Busch}
\begin{equation}
\label{En2D_ph}
- \ln \left( \frac{2 a_\mrm{2D}^2}{d_{\perp}^2}\right) = \psi\left(\frac{E_0-E}{2 \hbar \omega_{\perp}}\right),
\end{equation}
where $\psi(z)$ denotes the digamma function: $\psi(z) = (d/dz) \ln \Gamma (z)$. Expressing Eq. (\ref{En2D_ph}) 
in terms of dimensionless units related with $\omega_z$, we obtain 
\begin{equation}
\label{En2D}
- \ln \left( 2 a_\mrm{2D}^2 \eta \right) = \psi\left(-{\cal E}/(2\eta)\right).
\end{equation}
To find the connection between the two-dimensional and the quasi-two-dimensional energy spectrum we use the 
following approximate formula, valid for $\eta \ll 1$ (see Appendix~\ref{App:DetInt} for derivation):
\begin{equation}
\label{SumG}
{\cal F}(x) \approx - \Phi(x) - \ln \eta - \psi(x/\eta), \quad \eta \ll 1
\end{equation}
Substituting (\ref{SumG}) into (\ref{Energ1}), we obtain the condition for eigenenergies of excited states
in the quasi-two-dimensional traps:
\begin{equation}
\label{EnQ2D}
\frac{\sqrt{\pi}}{a} = \ln(\eta) + \Phi(-{\cal E}/2) + \psi\left(-{\cal E}/(2\eta)\right)
\end{equation} 
For the lowest excited states ($|{\cal E}|\ll 1$), we can neglect the energy dependence of 
$\Phi(-{\cal E}/2)$. In this regime it is straightforward to observe that Eqs. (\ref{En2D}) 
and (\ref{EnQ2D}) predict the identical energy spectrum, 
provided that the two-dimensional scattering length $a_{2D}$ is related with $a$ by 
\begin{equation}
a_\mrm{2D} = \frac{1}{\sqrt{2}} \exp\left( \frac{\Phi(0)}{2} - \frac{\sqrt{\pi}}{2 a} \right)
\end{equation}
Consequently, the two-dimensional coupling constant $g_\mrm{2D} = - \pi \hbar^2/(\mu\ln(k a_\mrm{2D}))$ 
expressed in the physical units is given by 
\begin{equation}
\label{g2D}
g_\mrm{2D} = \frac{2 \pi \hbar^2}{\mu} \frac{1}{ \sqrt{\pi}d/a - \Phi(0) - \ln(k^2d^2/2)},
\end{equation}
which agrees \footnote{In the coupling constant of Ref.~\cite{Petrov1}, instead of $\Phi(0) \approx 1.938$ 
a slightly different constant $\log(2 \pi) \approx 1.838$ appears. Ref.~\cite{Petrov2} provides
an accurate value of the numerical constant.}
with the results of Ref.~\cite{Petrov1,Petrov2}.

\begin{figure}
	 \includegraphics[width=8.5cm,clip]{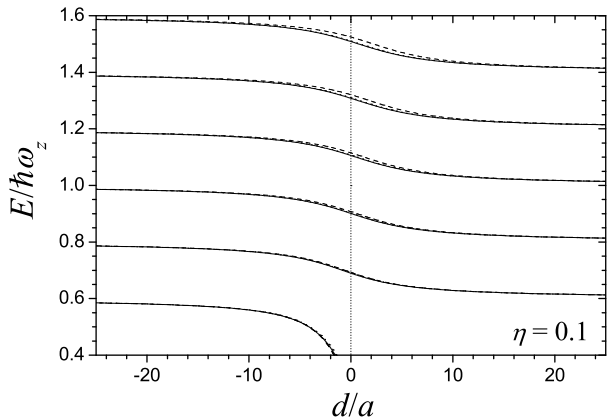}
	 \caption{
	 \label{Fig:E01}
	 Eigenenergies of the relative motion for two atoms interacting via $s$-wave pseudopotential and confined
	 in a harmonic trap with $\eta = \omega_{\perp}/\omega_z = 0.1$.
	 The exact energy levels (solid lines) are compared with 
	 predictions of the one-dimensional model with the energy-dependent 
	 (dotted line, indistinguishable from the solid one) and with the standard, energy-independent
	 renormalized scattering length (dashed lines).
	 The three-dimensional scattering length $a$ is scaled by the 
	 harmonic oscillator length $d = \sqrt{\hbar / (\mu \omega_z)}$.
	 }
\end{figure}

Similarly as for the quasi-one-dimensional traps, we have also verified the limits of applicability
of the two-dimensional model with renormalized scattering length. It turns out that the latter approach is
quite accurate 
up to $\eta \sim 1/10$. We note that in the case of $\eta =1/10$ the ratio between the harmonic oscillator
lengths in the radial and axial direction is equal to $\sqrt{10}$, therefore we would expect that 
the shape of the wave function is rather far from the quasi-two-dimensional one. Hence, the good agreement between
the exact energy levels and the two-dimensional spectrum is quite surprising. 
This feature is shown in Fig.~\ref{Fig:E01}. We observe 
that the two-dimensional approximation gives less accurate results for higher excited states 
around the unitarity point ($1/a=0$). This again can be improved by introducing an
energy-dependent two-dimensional scattering length 
\begin{equation}
a_\mrm{2D}(E) = \frac{d}{\sqrt{2}} \exp\left[ \frac{\Phi\left[(E_0-E)/(2 \hbar \omega_z)\right]}{2} - \frac{\sqrt{\pi} d}{2 a} \right],
\end{equation}
which can be easily derived by comparing \eqref{EnQ2D} with \eqref{En2D}. Eigenenergies calculated assuming 
the energy-dependent scattering length $a_\mrm{2D}(E)$ are in excellent agreement with exact ones, which is 
illustrated in Fig.~\ref{Fig:E01}.

\section{Wave functions}
\label{Sec:Fun}

\subsection{Axially symmetric trap of arbitrary anisotropy}
\label{Sec:FunAn}

Let us turn now to the analysis of the wave functions.
For $m \neq 0$ the wave functions of noninteracting particles vanish at ${\mbf r} =0$, 
and as a consequence they are not modified by the presence of the zero-range potential.
The nontrivial $m=0$ wave functions are given by Eq. (\ref{DefPsiE}), 
or by the integral representation (\ref{PsiE}), valid for ${\cal E} < 0$,
where in the place of ${\cal E}$ 
one has to substitute eigenenergies calculated from Eq.~\eqref{Energ1}. 
In the following we derive analytic formulas for the wave functions which are simpler and more 
convenient for numerical calculations.

We start from Eq. (\ref{DefPsiE}):
\begin{equation}
\label{PsiE1}
\Psi_{\cal E}(\mbf{r}) = \frac{\eta e^{-(\eta \rho^2+ z^2)/2}}{\pi^{3/2}} \sum_{m,k=0}^{\infty} 
\frac{H_k(0) H_k(z) L_m(\eta \rho^2)}{2^k k! (2 \eta m + k - {\cal E})},
\end{equation}
where we have substituted explicitly harmonic-oscillator wave functions. In the next step we rewrite the factor 
$(2 \eta m + k - {\cal E})$ in the denominator of (\ref{PsiE1}) as the integral \eqref{Int}. Then, we perform 
the summation 
only over a single variable $k$ or $m$, using the generating functions \eqref{SumL} and \eqref{SumH}. This yields an integral, which can be calculated analytically, and
the final result can be expressed in terms of a series in a single variable. 

Let us first perform the summation over the quantum 
number $k$. Applying the generating function (\ref{SumH}) we get 
\begin{align}
\nonumber
\Psi_{\cal E}(\mbf{r}) = & \frac{\eta e^{-(\eta \rho^2+ z^2)/2}}{\pi^{3/2}} \sum_{m=0}^{\infty} 
\left[  L_m(\eta \rho^2) \right. \\
\label{PsiE2}
& \times \left.
\int_0^{\infty} \! \! \! \ud t \ \frac{e^{-t (2 \eta m -{\cal E})}}{\sqrt{1-e^{-2t}}} 
\exp\left(- z^2 \frac{e^{-2t}}{1-e^{-2t}}\right)
\right].
\end{align}
The integration can be done analytically by introducing a new variable of integration $x= e^{-2t}/(1-e^{-2t})$
\cite{Gradshteyn}. In this way we find
\begin{align}
\nonumber
\Psi_{\cal E}(\mbf{r}) = & \frac{\eta}{2 \pi^{3/2} 2^{{\cal E}/2}} e^{-\eta \rho^2/2} \\
\label{PsiExp1}
& \times \sum_{m=0}^{\infty} 2 ^{\eta m} L_m(\eta \rho^2) \Gamma({\tst \frac{2 \eta m -{\cal E}}{2}}) 
D_{{\cal E}-2 \eta m}(|z|\sqrt{2}),
\end{align}
where $D_{\nu}(x)$ denotes the parabolic cylinder function. 
Since the parabolic cylinder functions $D_{\nu}(x)$ 
are well defined both for positive and negative values of index $\nu$, the validity of the result (\ref{PsiExp1})
is automatically extended to all values of energy ${\cal E}$. 
We note that parabolic cylinder functions $D_{\nu}(|z|\sqrt{2})$
are one-dimensional wave functions for two interacting atoms in a harmonic trap.
In the transverse direction the expansion (\ref{PsiExp1}) involves harmonic-oscillator wave functions.
The latter constitute an orthonormal basis, which simplifies analytic calculations of the 
matrix elements involving the wave functions in position representation.
As an example let us calculate the normalization factor ${\cal N}$
for the wave function $\Psi_{\cal E}(\mbf{r})$ : 
${\cal N}^{-2} = \int \ud^3 r |\Psi_{\cal E}(\mbf{r})|^2$, where for 
$\Psi_{\cal E}(\mbf{r})$ we substitute the expansion (\ref{PsiExp1}).
Integration over $\rho$ is trivial due to 
the orthonormal properties of the transverse wave functions,
whereas the axial integration involving $D_{\nu}(|z|\sqrt{2})^2$ can be performed analytically. This results in
\begin{align}
\label{N1}
{\cal N}^{-2} = \frac{\eta}{4 \pi} \sum_{m=0}^{\infty} \frac{\Gamma(-\frac{{\cal E}}{2}+\eta m)}
{\Gamma(-\frac{{\cal E}}{2}+\eta m+\frac{1}{2})}
\beta(-{\cal E} + 2 \eta m),
\end{align}
with $\beta(x)=[\psi((x+1)/2)-\psi(x/2)]/2$.

Another expansion is obtained when in Eq. (\ref{PsiE1}) we perform 
a summation over the quantum number $m$. Representing the denominator
of (\ref{PsiE1}) as the integral (\ref{Int}) and utilizing the summation formula (\ref{SumL}), we arrive at 
\begin{align}
\nonumber
\Psi_{\cal E}(\mbf{r}) = & \frac{\eta e^{-(\eta \rho^2+ z^2)/2}}{\pi^{3/2}} \sum_{k=0}^{\infty} 
\left[  \frac{H_k(0) H_k(z)}{2^k k!} \right. \\
\label{PsiE3}
& \times \left.
\int_0^{\infty} \! \! \! \ud t \ \frac{e^{-t (k -{\cal E})}}{1-e^{-2 \eta t}} 
\exp\left(- \eta \rho^2 \frac{e^{-2\eta t}}{1-e^{-2\eta t}}\right)
\right].
\end{align} 
The result of the integration 
can be expressed in terms of the confluent hypergeometric function $U(a,b,z)$, which yields
\begin{align}
\nonumber
\Psi_{\cal E}(\mbf{r}) = & \frac{e^{-(\eta \rho^2+ z^2)/2}}{2 \pi^{3/2}} \sum_{k=0}^{\infty} 
\left[ \frac{(-1)^k H_{2k}(z)}{2^{2k} k!} \right. \\
\label{PsiExp2}
& \times \left.
\Gamma({\tst \frac{k}{\eta} -\frac{{\cal E}}{2 \eta}}) U({\tst \frac{k}{\eta} -\frac{{\cal E}}{2 \eta}},1,\eta \rho^2).
\right],
\end{align} 
where we have substituted the values of the Hermite polynomials at $z=0$: $H_{2k}(0)= (-2)^k (2k-1)!!$, and 
$H_{2k+1}(0) = 0$. As in the previous case, this expansion can be utilized both for positive 
and negative values of energy ${\cal E}$, since the analytic continuation is provided automatically 
by the properties of $\Gamma(a)$ and $U(a,1,x)$ for negative values of the parameter $a$. 
As it can be easily verified,
$U(a,1,\eta \rho^2)$ is a wave function of the two interacting atoms in a two dimensional trap. 
In the axial directions, the expansion is done in the basis of one-dimensional harmonic oscillator wave functions.
We note that $U(a,1,\eta \rho^2)$ exhibits logarithmic divergence at
$\rho = 0$ and the result (\ref{PsiExp2}) is not correct for $\rho = 0$. In this
particular case the expansion (\ref{PsiExp1}) should be used instead. 

The expansion (\ref{PsiExp2}) can be used to obtain the normalization factor ${\cal N}$. Substituting (\ref{PsiExp2})
into ${\cal N}^{-2} = \int \ud^3 r |\Psi_{\cal E}(\mbf{r})|^2$ and performing first the integration in the 
axial direction, and then applying 
\begin{equation}
\int_0^{\infty} e^{-x} \ud x \ \Gamma(a)^2 U(a,1,x)^2 = \zeta_H(2,a)
\end{equation}
for the transverse integration we arrive at 
\begin{align}
\label{N2}
{\cal N}^{-2} = \frac{1}{4 \pi^{3/2} \eta} \sum_{m=0}^{\infty} \frac{(2m)!}{\left( 2^{m} m!\right)^2} 
\zeta_H\left(2,\frac{m}{\eta}-\frac{{\cal E}}{2\eta}\right).
\end{align}

In several applications of our theory, like evaluation of the exact dynamics for two interacting atoms, it is 
necessary to determine the complete set of eigenfunctions. The eigenfunctions that are modified 
by the interaction are given by representations \eqref{PsiExp1} and \eqref{PsiExp2}, with the energy ${\cal E}$ 
evaluated from the implicit equation \eqref{Energ1}. The remaining eigenfunctions are the harmonic-oscillator
wave functions that vanish at $\mbf{r}=0$. Applying the notation of Section \ref{Sec:SchrEq} they can be written 
as $\Phi_{n,m}(\rho,\varphi) \Theta_{k}(z)$ with $m>0$ or $m=0$ and $k$ odd. When $\omega_z$ and $\omega_\perp$
are incommensurate all the rest of the eigenstates is generated by \eqref{Energ1}. In the case of 
commensurable trapping frequencies, like for example in the particular case of $\eta=n$ or $\eta=1/n$, 
there can appear accidental degeneracies in the energy spectrum of the harmonic oscillator. Assuming that there
are $N$ degenerate eigenstates, the condition \eqref{Energ1} generates only the single state affected by the
interaction, while the remaining $N-1$ states can be determined, for instance from the Gram-Schmidt orthonormalization
procedure, and they will automatically vanish at $\mbf{r}=0$.

\subsection{Quasi-1D regime}
\label{Sec:FunQ1D}

Now we discuss the behavior of the wave functions in the limit of very elongated traps: $\eta \gg 1$. 
We start from the integral representation (\ref{PsiE}) to obtain the wave function of the ground state.
For $\eta \gg 1$ and $|{\cal E}| \sim \eta$  
the main contribution to the integral comes from the region of small $t$.
Expansion of the integrand to the lowest order in
$t$ yields
\begin{align}
\label{PsiQ1D} 
\Psi_{\cal E}(\mbf{r}) \approx & \frac{\eta}{(2 \pi)^{\frac{3}{2}}}
\int_{0}^{\infty} \! \! \! \! \ud t \ \frac{\exp\!\left[ t E
-\frac{z^2}{2t} - \frac{\eta \rho^2}{2} \coth(\eta t)
\right]} {\sqrt{t} \sinh (\eta t)}.
\end{align}
Changing the variable of integration and expressing the final result in the physical units we obtain
\begin{align}
\label{PsiQ1D_1} 
\Psi_{\cal E}(\mbf{r}) \approx & \frac{\sqrt{\eta}}{(2 \pi)^{\frac{3}{2}}}
\int_{0}^{\infty} \! \! \! \! \ud t \ \frac{\exp\!\left[ \frac{t E}{\hbar \omega_{\perp}}
-\frac{z^2}{2t d_{\perp}} - \frac{\rho^2}{2 d_{\perp}} \coth(\eta t)
\right]} {\sqrt{t} \sinh (t)}.
\end{align}
We note that the wave function of a bound state depends only on the trapping frequency $\omega_{\perp}$ 
in the tightly confined direction \footnote{The dependence on $\eta$ in the prefactor of \eqref{PsiQ1D_1} is 
eliminated by normalizing $\Psi_{\cal E}(\mbf{r})$.}. This feature has been already observed on the level of 
eigenenergies.

The wave function of the bound state in quasi-one-dimensional traps can be also calculated from the following expression:
\begin{equation}
\Psi_{\cal E} (\mbf{r}) 
\approx \frac{\eta e^{-\eta \rho^2/2}}{2 \pi} \sum_{m=0}^{\infty}
L_m(\eta \rho^2)
\frac{\exp \left(-2 |z|\sqrt{-\frac{{\cal E}}{2}+m \eta}\,\right)}{\sqrt{-\frac{{\cal E}}{2}+ m \eta}}
\label{PsiQ1D_2}
\end{equation}
To derive \eqref{PsiQ1D_2} we 
expand \eqref{PsiE2} for small $t$ and perform analytically the integration over $t$.
From Eq.~\eqref{PsiQ1D_2} one can easily derive approximate 
axial and radial profiles of the bound-state wave function \cite{Idziaszek}, 
which agrees very well with the exact one evaluated from Eqs.~\eqref{PsiExp1} and \eqref{PsiExp2}.

\begin{figure}
	 \includegraphics[width=8cm,clip]{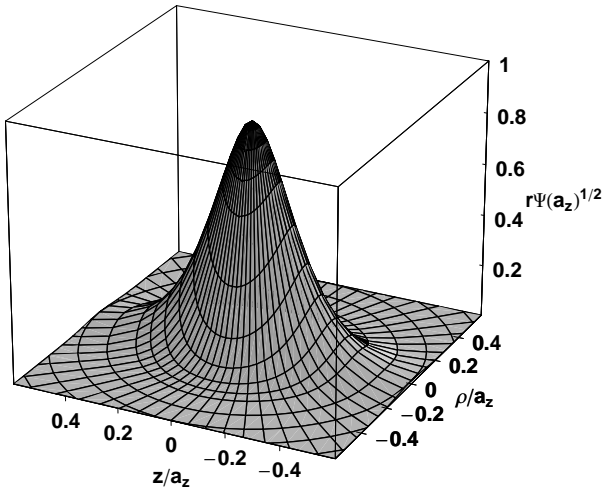}
	 \caption{
	 \label{fig:3D100_0}
	 Exact wave function $r \Psi({\mbf r})$ for two atoms interacting via $s$-wave pseudopotential 
	 and trapped in a harmonic potential with $\eta = \omega_{\perp}/\omega_z = 100$. The figure shows
	 the ground state for the scattering length $a=\pm \infty$. All lengths are scaled the units of  
	 $a_z = \sqrt{\hbar / (\mu \omega_z)}$.
	 }
\end{figure}

The exact ground-state wave function for $a=\pm \infty$ and $\eta=100$ is presented in Fig.~\ref{fig:3D100_0}.
For small $r$ it is almost isotropic, due to the presence of the divergent factor $1/(2 \pi r)$, while 
for larger values of $r$ the wave function is slightly elongated in the $z$-direction, which reflects 
the geometry of the trap. 

Let us discuss now the properties of the excited-state wave functions. In the regime of
energies corresponding to the lowest excited states and for the arguments $z$ not too small, it turns out that 
the first term of the series (\ref{PsiExp1}) dominates the sum. This is a consequence of the
behavior of the parabolic cylinder function $D_{\nu}(x)$, which are fast decaying when 
index $\nu$ becomes large and negative. Thus the approximate wave function of the excited state reads
\begin{align}
\label{PsiQ1DEx}
\Psi_{\cal E}(\mbf{r}) \approx \frac{\eta}{2 \pi^{3/2} 2^{{\cal E}/2}} e^{-\eta \rho^2/2} 
\Gamma\left({\tst -\frac{{\cal E}}{2}}\right) D_{{\cal E}}\left(|z|\sqrt{2}\right).
\end{align}
Obviously, the latter approximation is not valid for small $r$, where the exact wave function exhibits a divergent 
behavior: $1/(2 \pi r)$. Fortunately, the main part of the wave function is located for larger values of $r$, and
the region of small $r$ gives a rather small contribution to the total wave function. On the other hand, we should keep in mind that the delta pseudopotential is only an approximation, and for small $r$, 
comparable to the effective range of the physical potential, 
the behavior of the real wave function is quite different from the predictions based on the
pseudopotential. 

To estimate the accuracy of the approximation (\ref{PsiQ1DEx}) one can consider the contribution
of the first term in the series (\ref{N1}), 
which comes from the integral involving the square of the wave function (\ref{PsiQ1DEx}).
For the range of energies corresponding to the first ten excited states, and for $\eta=100$
we obtain that 
the first term accounts for more than 0.9986 of the whole sum in Eq.~\eqref{N1}.

\begin{figure}
	 \includegraphics[width=8.5cm,clip]{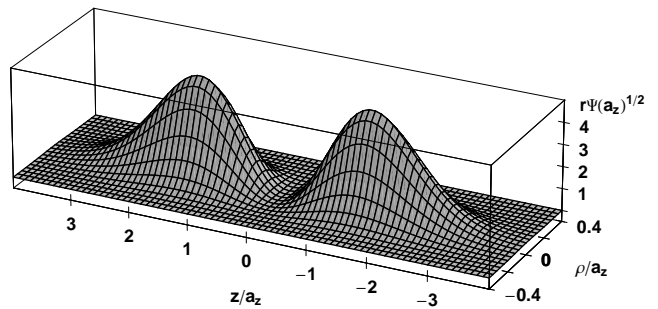}
	 \caption{
	 \label{fig:3D100_1}
	 Exact wave function $r \Psi({\mbf r})$ for two atoms interacting via $s$-wave pseudopotential 
	 and trapped in a harmonic potential with $\eta = \omega_{\perp}/\omega_z = 100$. The figure presents 
	 the first excited state for the scattering length $a=\pm \infty$. All lengths are scaled in units of  
	 $a_z = \sqrt{\hbar / \mu \omega_z}$.
	 }
\end{figure}

\begin{figure}
	 \includegraphics[width=8.5cm,clip]{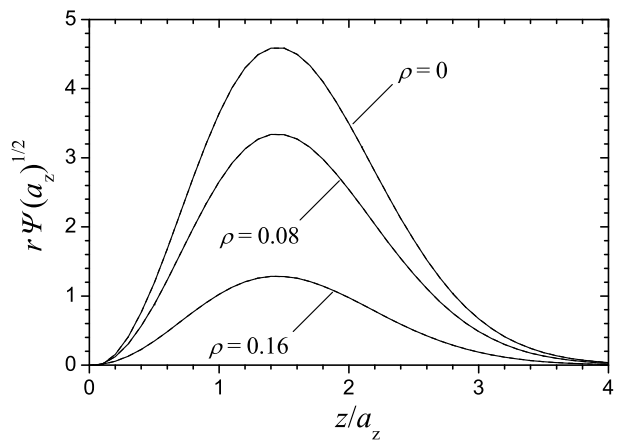}
	 \caption{
	 \label{fig:Cut100_1z}
	 The axial profiles of the wave function presented in Fig.~\ref{fig:3D100_1}, evaluated for 
	 $\rho=0$, $\rho=0.08$ and $\rho=0.16$.	The exact profiles (solid lines) are compared with the predictions 
	 of the quasi-one-dimensional approximation given by Eq. (\ref{PsiQ1DEx}).
	 All lengths are scaled in units of $a_z = \sqrt{\hbar / (\mu \omega_z)}$.
	 }
\end{figure}

\begin{figure}
	 \includegraphics[width=8.5cm,clip]{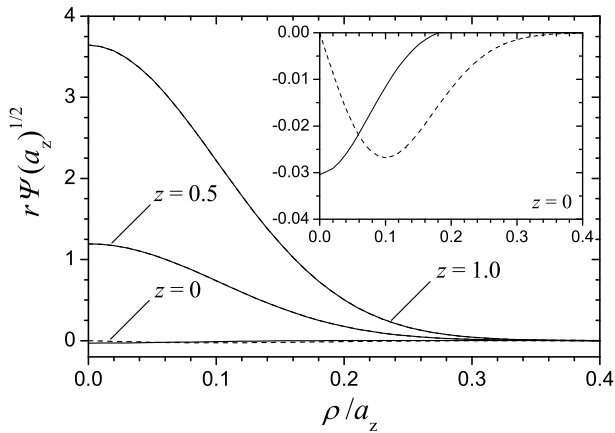}
	 \caption{
	 \label{fig:Cut100_1r}
	 The radial profiles of the wave function presented in Fig.~\ref{fig:3D100_1}, evaluated for 
	 $z=0$, $z=0.5$ and $z=1.0$.	The exact profiles (solid lines) are compared with the predictions 
	 of the quasi-one-dimensional approximation given by Eq. (\ref{PsiQ1DEx}). 
	 The inset shows the details of $z=0$ profile for small values of $\rho$.
	 All lengths are scaled in units of $a_z = \sqrt{\hbar / (\mu \omega_z)}$.
	 }
\end{figure}

Fig.~\ref{fig:3D100_1} shows the exact wave function of the first excited state for $a=\pm \infty$ and $\eta=100$. 
The wave function was evaluated from expansions (\ref{PsiExp1}) and (\ref{PsiExp2}).
We observe that the wave function is strongly elongated in the axial
direction and that the region of small $r$ gives rather small contribution to the 
total wave function. In the transverse direction the wave function exhibits the exponential 
behavior predicted by Eq. (\ref{PsiQ1DEx}). 
These matters are illustrated in more details in Figs.~\ref{fig:Cut100_1z} 
and \ref{fig:Cut100_1r} showing, respectively, the axial and the transverse profiles of the wave function. 
The figures compare the exact profiles with the quasi-one-dimensional prediction of Eq. (\ref{PsiQ1DEx}).
The approximate curves fit very well the exact wave function, except for the transverse profile calculated for $z=0$.
In this case, the small $\rho$ behavior of the wave function, presented in more detail in the inset of 
Fig.~\ref{fig:Cut100_1z}, is dominated by the divergent term $1/(2 \pi r)$, which
is absent in the approximation (\ref{PsiQ1DEx}).

\subsection{Quasi-2D regime}
\label{Sec:FunQ2D}

In this section we analyze the properties of the wave functions in the regime $\eta \ll 1$. 
Let us first focus on the wave function of the ground state. In this case we apply the  
integral representation (\ref{PsiE}), where for $|{\cal E}| \sim 1$ and $\eta \ll 1$ the main 
contribution to the integration comes from $t \sim 1$. Expanding the integrand for small $\eta$ and expressing the 
result in the physical units we obtain
\begin{align}
\label{PsiQ2D_1} 
\Psi_{\cal E}(\mbf{r}) \approx & \frac{1}{(2 \pi)^{\frac{3}{2}}}
\int_{0}^{\infty} \! \! \! \! \ud t \ \frac{\exp\!\left[ t E
-\frac{z^2}{2 d} \coth t - \frac{\eta \rho^2}{2 d t}
\right]} {\sqrt{\sinh(t)} t}.
\end{align}
We note that the ground-state wave function depends exclusively on the trapping frequency in the $z$ direction,
thus its size is given approximately by $d = \sqrt{\hbar/(\mu \omega)}$

As for the quasi-one-dimensional traps, we have also found another representation for the wave function of
the ground-state. It can be applied, for instance, to derive approximate axial and radial profiles, 
which agree very well with the exact wave functions, as we have shown in \cite{Idziaszek}. Expanding \eqref{PsiE3} for small $\eta$ and performing an integration over $t$ we arrive at 
\begin{equation}
\Psi_{\cal E}(\rho,z=0) \approx \frac{e^{-z^2/2}}{\pi^{3/2}} \sum_{k=0}^{\infty} 
\frac{ H_k(0) H_k(z)}{2^k k!} 
K_0 \left({\tst \rho \sqrt{2k -2{\cal E}}} \right),
\label{PsiQ2D_2}
\end{equation}
where $K_0(x)$ is a modified Bessel function.

\begin{figure}
	 \includegraphics[width=8cm,clip]{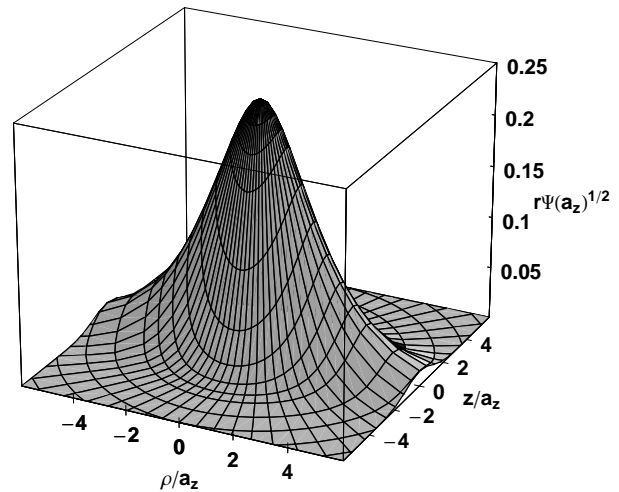}
	 \caption{
	 \label{fig:3D001_0}
	 Exact wave function $r \Psi({\mbf r})$ for two atoms interacting via $s$-wave pseudopotential 
	 and trapped in a harmonic potential with $\eta = \omega_{\perp}/\omega_z = 0.01$. The figure shows
	 the ground state for the scattering length $a=\pm \infty$. All lengths are scaled in units of  
	 $a_z = \sqrt{\hbar /(\mu \omega_z)}$.
	 }
\end{figure}

Fig.~\ref{fig:3D001_0} presents the exact ground-state wave function for $a=\pm \infty$ and $\eta=0.01$, 
evaluated from Eqs.~\eqref{PsiExp1} and \eqref{PsiExp2}. We note 
that the ground-state wave function for small $r$ is nearly isotropic, while for larger $r$ it is slightly 
elongated in the direction of weaker confinement. The anisotropy of a bound-state for $\eta \ll 1$
seems to be larger than in quasi-one-dimensional traps. 

Let us investigate now the properties of the wave functions for excited states. 
The approximate form of the wave function can be found from the expansion (\ref{PsiExp2}). 
In the regime of energies corresponding to the lowest
excited states, the sum in Eq.~(\ref{PsiExp2}) is dominated by the first term. 
This results from the asymptotic properties of 
the confluent hypergeometric function $U(c,1,x)$, which decays faster in $x$
for larger values of the parameter $c$. Hence, 
the approximate wave function of excited states in quasi-two-dimensional traps is given by 
\begin{align}
\label{PsiQ2DEx}
\Psi_{\cal E}(\mbf{r}) \approx \frac{e^{-(\eta \rho^2+z^2)/2}}{2 \pi^{3/2}}   
\Gamma({\tst -\frac{{\cal E}}{2 \eta}}) U({\tst -\frac{{\cal E}}{2 \eta}},1,\eta \rho^2).
\end{align}
Approximate result~(\ref{PsiQ2DEx}) cannot be directly used for $\rho=0$, where the confluent 
hypergeometric function $U$ exhibits logarithmic divergence. Due to the same reason,
the latter approximation is also not valid 
when $r \to 0$, where the exact wave function behaves as $1/(2 \pi r)$. 
Similarly as for the quasi-one-dimensional wave function,  
the leading part of 
the wave function is located outside the region of small $\rho$ and $r$, which makes the approximation 
(\ref{PsiQ2DEx}) quite accurate. 

To estimate the quality of the approximation (\ref{PsiQ2DEx}) we considered the series (\ref{N2}),
where the first term comes from the integral of square modulus of \eqref{PsiQ2DEx}.
Within the range of energies corresponding to the first ten excited states, and for $\eta=0.01$,
we obtain that the first term of (\ref{N2}) contributes to more than $0.9985$ of the total sum.

\begin{figure}
	 \includegraphics[width=8.5cm,clip]{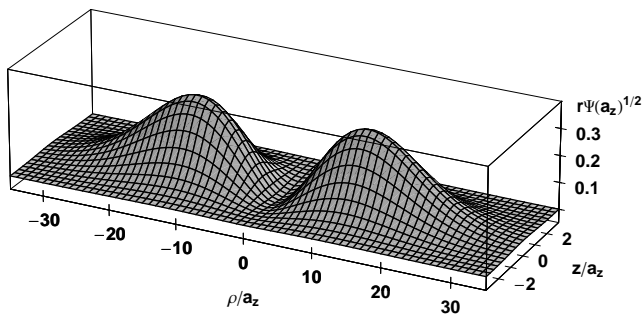}
	 \caption{
	 \label{fig:3D001_1}
	 Exact wave function $r \Psi({\mbf r})$ for two atoms interacting via $s$-wave pseudopotential 
	 and trapped in a harmonic potential with $\eta = \omega_{\perp}/\omega_z = 0.01$ The figure presents 
	 the first excited state for the scattering length $a=\pm \infty$. All lengths are scaled in units of  
	 $a_z = \sqrt{\hbar / (\mu \omega_z)}$.
	 }
\end{figure}

\begin{figure}
	 \includegraphics[width=8.5cm,clip]{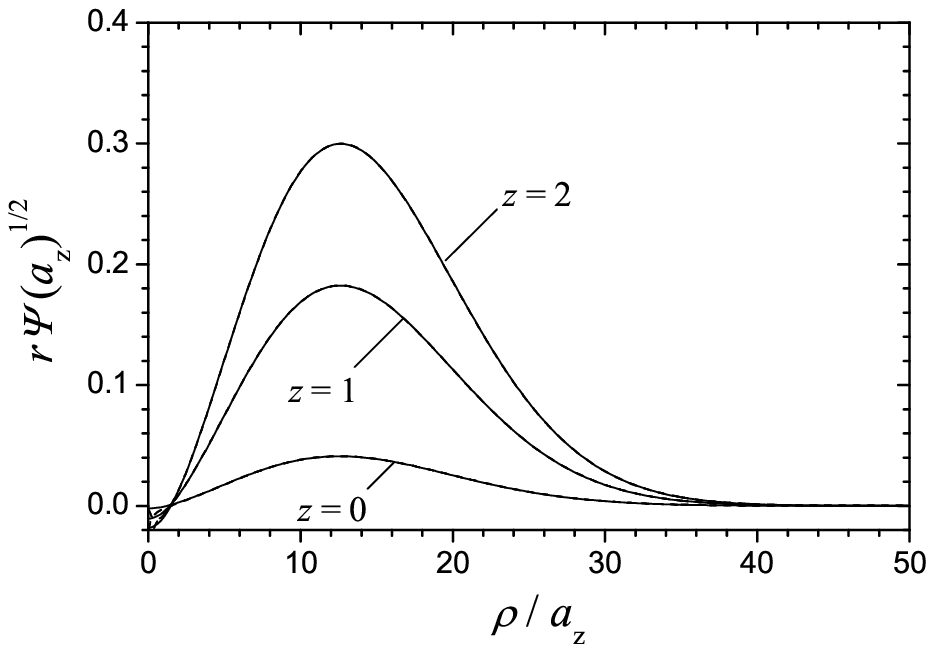}
	 \caption{
	 \label{fig:Cut001_1r}
	 The radial profiles of the wave function presented in Fig.~\ref{fig:3D001_1}, evaluated for 
	 $z=0$, $z=1$ and $z=2$. The exact profiles (solid lines) are compared with predictions 
	 of the quasi-two-dimensional approximation (\ref{PsiQ2DEx}). 
	 All lengths are scaled in units of $a_z = \sqrt{\hbar /(\mu \omega_z)}$.
	 }
\end{figure}

\begin{figure}
	 \includegraphics[width=8.5cm,clip]{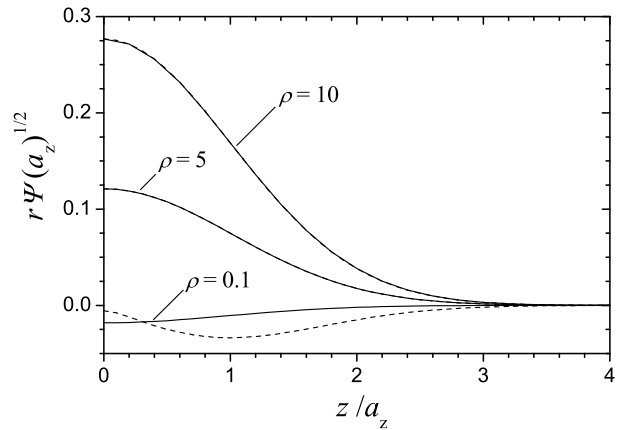}
	 \caption{
	 \label{fig:Cut001_1z}
	 The axial profiles of the wave function presented in Fig.~\ref{fig:3D001_1}, evaluated for 
	 $\rho=0.1$, $\rho=5$ and $\rho=10$. The exact profiles (solid lines) are compared with predictions 
	 of the quasi-two-dimensional approximation (\ref{PsiQ2DEx}).
	 All lengths are scaled in units of $a_z = \sqrt{\hbar /(\mu \omega_z)}$.
	 }
\end{figure}

The exact wave function of the first excited state for $a=\pm \infty$ and $\eta=0.01$ is presented in 
Fig.~\ref{fig:3D001_1}. It is evaluated from the expansions (\ref{PsiExp1}) and (\ref{PsiExp2}).
We observe that the wave function of the excited state is highly anisotropic,
and elongated in the radial direction, reflecting the geometry of the trap. More detailed behavior can be deduced 
from Figs.~\ref{fig:Cut001_1r} and \ref{fig:Cut001_1z}, showing respectively the transverse and the axial profiles 
of the wave function. These figures compare the exact profiles with the quasi-two-dimensional approximation
given by Eq. (\ref{PsiQ2DEx}). In the case of transverse
profiles all the approximate curves are almost indistinguishable 
from the exact ones, except in the region of small $\rho$ when the approximation (\ref{PsiQ2DEx})
ceases to be valid due to the logarithmic divergence of $U$. 
On the other hand, the approximate axial profiles for $\rho=5$ and $\rho=10$ fit very well
the exact wave function, while for $\rho=0.1$ the approximation (\ref{PsiQ2DEx}) clearly deviates
from the exact result. The latter discrepancy can be again attributed to the logarithmic divergence 
of (\ref{PsiQ2DEx}) at small $\rho$.

\section{Feshbach resonances}
\label{Sec:Feshbach}

In this section we extend our theory to the system of two interacting atoms in 
the presence of Feshbach resonances. The latter technique is widely used in 
recent experiments on ultracold atoms and allows to tune the value of the scattering length
by changing the strength of applied magnetic field. 
Close to the Feshbach resonance, the scattering length can take a very
large value or can change its sign. Obviously, when $a$ becomes too large, the interaction between two atoms cannot 
be described in terms of the standard pseudopotential (\ref{Vi}). To see this we recall that (\ref{Vi})
is obtained from the more general form of the pseudopotential \cite{HuangPs}
\begin{equation}
\label{Vps}
V (\mathbf{r}) = - \frac{2 \pi \hbar^2 \tan \delta_0(k)}{\mu k} \delta(\mathbf{r}) \frac{\partial}{\partial r} r,
\end{equation}
in the limit when $\tan \delta_0 \approx - k a$. As it can be easily verified, the latter assumption is fulfilled
for $k a \ll 1/(k R_0)$ where $R_0$ is the effective range, whereas application of the zero-range pseudopotential
requires $k R_0 \ll 1$.

Following the work of Bolda {\it et al.} \cite{Bolda}, we introduce an effective 
scattering length, which is defined by 
\begin{equation}
\label{aEff}
a_{\mathrm{eff}}(E) = - \frac{\tan \delta_0 (k)}{k}
\end{equation}
with $k$ related to the kinetic energy by $E = \hbar^2 k^2 /(2 \mu)$. 
In general, the effective scattering length is energy 
dependent, however, for sufficiently small $k$, $a_{\mathrm{eff}}(E)$ reduces to the standard scattering length.
Now we can reformulate our theory, replacing the standard scattering length $a$ by $a_{\mathrm{eff}}(E)$.
In this way Eq. (\ref{Energ1}) determining the eigenenergies of the system is substituted by
\begin{equation}
\label{EnergSelf}
- \frac{\sqrt{2 \pi}}{a_{\mathrm{eff}}(E)} = {\cal F}\left(\frac{E_0-E}{2}\right).
\end{equation}
By solving the latter equation in a self-consistent way we obtain the eigenenergies, 
which are valid for arbitrarily large values of the scattering length. Moreover, by performing the analytic continuation
of \eqref{aEff} to negative energies (imaginary $k$), one can properly account for the whole spectrum of bound states
\cite{Stock}.

We stress that the validity of the
discussed model is based on the assumption that the effective range of the physical potential is much smaller than 
the harmonic oscillator length $d$. In this regime it is justified to use the pseudopotential (\ref{Vps}),
which was derived for free space. This assumption also guarantees that at distances comparable to the range
of the physical potential, the kinetic energy, which enters the pseudopotential (\ref{Vps}) through $k$,
is equal to the total energy $E$.

To apply the concept of the effective scattering length to Feshbach resonances,
we have to specify the dependence of the $s$-wave phase shift on $k$. 
The theory \cite{Moerdijk,Timmermans,Goral} predicts the following dependence for $\delta_0$
\begin{equation}
\label{delta0}
\delta_0 = \delta_{bg} - \arctan \left(\frac{\gamma}{E-E_m-\Delta_m}\right),
\end{equation}
where $\delta_{bg}$ is the background phase shift, $E_m$ denotes the energy of the resonance, 
and $\Delta_m$ is the energy shift due to the coupling between the open and closed channels. 
The parameter $\gamma$ is a "reduced width" of the resonance which is related to the usual width $\Gamma$
by $\Gamma = 2 \gamma k$ \cite{Timmermans}.
Close to the Feshbach resonance we can assume that the energy $E_m$ varies linearly with the magnetic field strength
\begin{equation}
E_m(B) = E_m^{\prime} (B - B_{\mrm{res}}),
\end{equation}
where 
\begin{equation}
E_m^{\prime} = \left.\frac{d E_m}{d B}\right|_{B_{\mrm{res}}},
\end{equation} 
and $B_{\mrm{res}}$ denotes the magnetic field strength at 
which the energy of the closed channel crosses the dissociation threshold of the open channel. By combining Eqs.
(\ref{aEff}) and (\ref{delta0}), after some straightforward algebra, we obtain the following result for the
energy-dependent scattering length
\begin{equation}
\label{aEff1}
a_{\mrm{eff}}(E) = a_{\mrm{bg}} \left[ 1 -\frac{ \Delta B \left(1+\frac{E}{E_b}\right)}{B - 
\left( B_0 + E/E_m^{\prime} - \Delta B \frac{E}{E_b} \right) } \right],
\end{equation}
where $E_b = \hbar^2 /(m a_{\mrm{bg}}^2)$ is the binding energy corresponding to the background scattering length. The 
resonance width $\Delta B$ and the resonance position $B_0$ are related with to the previous parameters by
\begin{equation}
B_0 = B_{\mrm{res}} - \frac{\Delta_m}{E_m^{\prime}},
\end{equation}
and
\begin{equation}
\Delta B = \frac{\gamma}{a_{\mrm{bg}} E_m^{\prime}}.
\end{equation}
For sufficiently small energies ($E \ll E_b,E_m^{\prime}B_0$) the effective scattering length becomes
independent of $E$, and Eq. (\ref{aEff1}) reduces to the well-known formula
\begin{equation}
\label{aEff2}
a_{\mrm{eff}}(E) = a_{\mrm{bg}} \left[ 1 -\frac{ \Delta B}{B - B_0} \right].
\end{equation}

\begin{figure}
	 \includegraphics[width=8.5cm,clip]{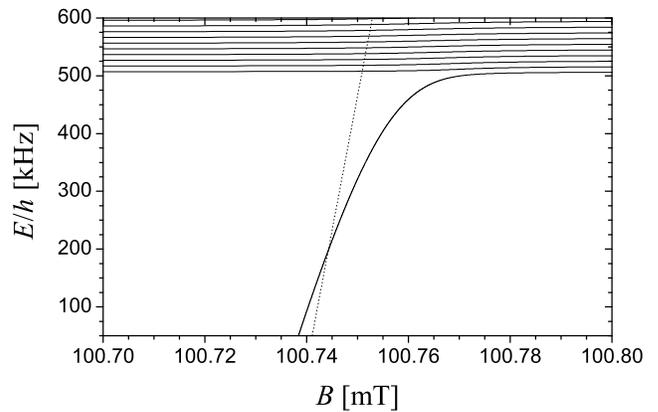}
	 \caption{
	 \label{fig:ERb100}
   Energy spectrum for two ${}^{87}$Rb atoms versus magnetic field $B$ near the Feshbach resonance 
   at $100$~mT. The atoms are confined in an axially symmetric trap with $\omega_z = 5$~kHz and  
   $\omega_{\perp} = 500$~kHz. The dotted line shows the value of the magnetic field at which the 
   energy-dependent scattering length $a_{\mrm{eff}}(E)$ diverges.
	 }
\end{figure}

\begin{figure}
	 \includegraphics[width=8.5cm,clip]{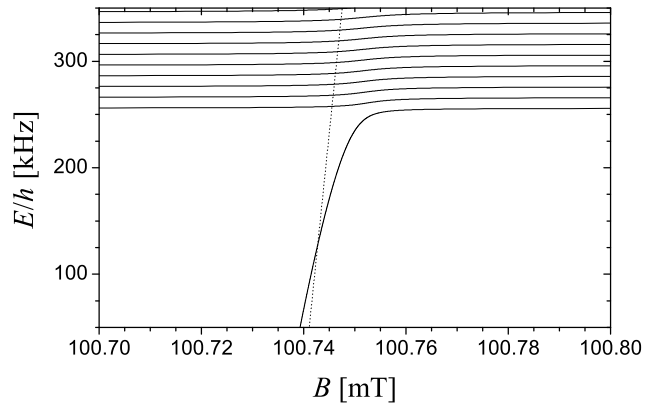}
	 \caption{
	 \label{fig:ERb001}
	 Energy spectrum for two ${}^{87}$Rb atoms versus magnetic field $B$ near a Feshbach resonance 
   at $100$~mT. The atoms are confined in an axially symmetric trap with $\omega_z = 500$~kHz and  
   $\omega_{\perp} = 5$~kHz. The dotted line shows the value of the magnetic field at which the 
   energy-dependent scattering length $a_{\mrm{eff}}(E)$ diverges.
 	 }
\end{figure}

Employing Eqs. (\ref{EnergSelf}) and (\ref{aEff1}) we have calculated the
energy spectrum for two ${}^{87}$Rb atoms interacting close to the Feshbach resonance at $100$~mT 
\footnote{For values of the physical parameters for ${}^{87}$Rb atoms at
the Feshbach resonance near $100$~mT see for example Ref. \cite{Goral}.}.
Fig.~\ref{fig:ERb100} presents energy levels versus magnetic field in the quasi-one-dimensional trap with 
$\eta =100$ and $\omega_z = 5$~kHz. 
In addition it shows the values of the magnetic field for which the effective
scattering length diverges. From the plot it is clear that in the considered range of energies the position of the 
resonance changes with the energy. A similar calculation performed for 
the quasi-two-dimensional trap with $\eta = 0.01$ and $\omega_z = 500$~kHz is shown on Fig.~\ref{fig:ERb001}.


\section{Conclusion}
\label{Sec:Conclusion}

In summary, we have presented a detailed analysis of the system of two 
interacting atoms confined in an axially symmetric harmonic trap.
We discussed in detail different regimes, in particular the quasi-one-
and the quasi-two-dimensional geometries. We have shown that these two regimes 
can be applied already for the traps with $\eta \gtrsim 10$ and $\eta \lesssim 0.1$, respectively.
In this way we demonstrate that, at least on the level of two-atom physics, realization of low-dimensional 
systems does not require extremely large (or small) ratios of the transverse to axial trapping frequencies.

We have applied our analytical results to studying the system of two atoms with interaction modified by 
a Feshbach resonance. To this end we have utilized the concept of an energy-dependent scattering length 
\cite{Bolda}, and employing well-known results from the theory of Feshbach resonances we have derived 
an explicit formula determining the energy spectrum in terms of the standard parameters describing 
the resonance.
Our results can be directly implemented to calculate exact dynamics of the two ultracold atoms in harmonic traps 
with arbitrary large trapping frequencies and in the presence of Feshbach resonances. This is 
particularly important in the context of implementation of  
quantum information processing in systems of trapped ultracold atoms.

\begin{acknowledgments}

The authors are grateful to L.P. Pitaevskii and G. Orso for valuable discussions.
We thank E. Bolda for making available his programs, which have helped us to verify the validity of our 
numerical calculations.
We acknowledge 
financial support from the European Union, contract number IST-2001-38863 (ACQP), the FP6-FET Integrated
Project CT-015714 (SCALA) and a EU Marie Curie Outgoing International Fellowship, and from the National 
Science Foundation through a grant for the Institute for Theoretical Atomic, Molecular and Optical Physics 
at Harvard University and Smithsonian Astrophysical Observatory. 

\end{acknowledgments}

\appendix

\section{Details of derivation for quasi-two-dimensional spectrum} 
\label{App:DetInt}

In this appendix we present the derivation of Eqs. (\ref{PhiSer}) and (\ref{SumG}), 
which we used in the analysis of energy spectrum in quasi-two-dimensional traps. 
Proof of the former formula starts from the integral defining the function $\Phi (x)$
\begin{equation}
\label{A1Phi}
- \Phi (x) - \ln x = \int_{0}^{1} \ud t \ B\left(x+t,{\tst -\frac{1}{2}}\right), \quad x>0,
\end{equation}
where $B(x,y) = \Gamma(x) \Gamma(y)/\Gamma(x+y)$ is the Euler Beta function.
Next, we apply the series representation for $B(x,1/2)$ \cite{Gradshteyn}
\begin{equation}
\label{A1BSer}
B(x,1/2) = \frac{1}{x}+ \sum_{k=1}^{\infty} \frac{(2k)!}{(2^k k!)^2} \frac{1}{x+k}.
\end{equation}
The series expansion of $B(x,-1/2)$ can be found by combining Eq. (\ref{A1BSer}) with the following identity, 
which follows directly from the definition of the Beta function
\begin{equation}
\label{A1Brel}
B(x,-1/2) = -2\left(x-\frac{1}{2}\right)B(x,1/2).
\end{equation}
Inserting the series expansion for $B(x,-1/2)$ into
the integral in Eq. (\ref{A1Phi}), and performing the integration term by term, we arrive at the final result 
\begin{align}
\label{A1PhiSer}
\Phi (x) = & 2 - \ln (1+x) \nonumber \\
& {} + 2 \sum_{k=1}^{\infty} \frac{(2k)!}{(2^k k!)^2} \left[(k+{\tst \frac{1}{2}})\ln\frac{x+k}{x+k+1}+1\right]
\end{align}

Now we turn to the derivation of Eq. (\ref{SumG}). We begin with the multiplication formula for the 
function $\psi(z)$ \cite{Gradshteyn} 
\begin{equation}
\psi(nz) = \frac{1}{n} \sum_{k=0}^{n-1} \psi\left(z + \frac{k}{n}\right) + \ln n,
\end{equation}
and use the definition of $B(x,y)$ to obtain the following identity:
\begin{align}
\nonumber
- \frac{2 \sqrt{\pi}}{n} & \sum_{m=0}^{n-1} 
\frac{\Gamma(x+\frac{m}{n})}{\Gamma(x-\frac{1}{2}+\frac{m}{n})} + \psi(n x) = \\
\label{A1SumG}
& \frac{1}{n} \sum_{m=0}^{n-1} \left(B\left({\tst x+\frac{m}{n},-\frac{1}{2}}\right) + \psi\left(
{\tst z + \frac{m}{n}}\right) \right)
+ \ln n.
\end{align}
For $n \gg 1$ we approximate the summation on the r.h.s. of Eq. (\ref{A1SumG}) by integration. 
Since the singularities of $B(x,-1/2)$ and $\psi(x)$ cancel each other at $x=0$, 
the replacement of the summation by integration is valid for $x > -1$, where the integrand 
is free from singularities. The latter approximation results in 
\begin{align}
\nonumber
-\frac{2 \sqrt{\pi}}{n} & \sum_{m=0}^{n-1}
\frac{\Gamma(x+\frac{m}{n})}{\Gamma(x-\frac{1}{2}+\frac{m}{n})} + \psi(n x) \stackrel{n\gg1}{\approx}  \\
\label{A1SumG1}
& \int_{0}^{1} \! \! \! \ud t \left( B\left(x+t,{\tst -\frac{1}{2}}\right) + \psi(x + t) \right)+ \ln n.
\end{align}
The integral of the first term can be expressed in terms of the function $\Phi(x)$ 
(cf. Eq. (\ref{A1Phi})), whereas the integration of
$\psi(x)$ is trivial and follows directly from the 
definition: $\psi(x) = (d/dx) \ln \Gamma (x)$. Finally we obtain 
\begin{align}
-\frac{2 \sqrt{\pi}}{n} \sum_{m=0}^{n-1}
\frac{\Gamma(x+\frac{m}{n})}{\Gamma(x-\frac{1}{2}+\frac{m}{n})} + \psi(n x) \stackrel{n\gg1}{\approx}
- \Phi(x) + \ln n.
\label{A1SumG2}
\end{align}
\mbox{}


\begin{thebibliography}{99}

\bibitem{Bloch} See for example: I. Bloch, Physics World {\bf 17}, 25 (2004), and references therein. 

\bibitem{microtraps} R. Folman {\em et al.}, Adv. At. Mol. Opt. Phys. {\bf 48} 263 (2002); R. Dumke {\em et al.}, Phys. Rev. Lett. {\bf 89}, 97903 (2002).

\bibitem{Grangier} N. Schlosser {\em et al}., Nature {\bf 411}, 1024 (2001).

\bibitem{BEC-BCS} C.A. Regal, M. Greiner and D.S. Jin, Phys. Rev. Lett.
{\bf 92}, 040403 (2004); M.W. Zwierlein {\it et al.}, Phys. Rev.
Lett. {\bf 92}, 120403 (2004); C. Chin {\it et al.}, Science {\bf 305},
1128 (2004); J. Kinast {\it et al.}, Science {\bf 307}, 1296 (2005);
G.B. Patridge {\it et al.}, Phys. Rev. Lett. {\bf 95}, 020404 (2005).

\bibitem{Mott} M. Greiner {\it et al.}, Nature {\bf 415} 39, (2002); 
T. St\"oferle, H. Moritz, C. Schori, M. K\"ohl, 
and T. Esslinger, Phys. Rev. Lett. {\bf 92}, 130403 (2004); K. Xu {\it et al.}, Phys. Rev. A 
{\bf 72}, 043604 (2005).

\bibitem{Busch} T. Busch, B.-G. Englert, K. Rz\c{a}\.{z}ewski, and M. Wilkens, Found. Phys. {\bf 28}, 549 (1998).

\bibitem{Idziaszek} Z. Idziaszek, and T. Calarco, Phys. Rev. A {\bf 71}, 050701(R) (2005).

\bibitem{Fermi} E. Fermi, Ricera Sci. {\bf 7}, 12 (1936).

\bibitem{HuangPs} K. Huang, and C.N. Yang, Phys. Rev. {\bf 105}, 767 (1957); K. Huang, {\it Statistical Mechanics} (John Willey \& Sons, New York, 1963).

\bibitem{Stock} R. Stock, A. Silberfarb, E.L. Bolda, and I.H. Deutsch, Phys. Rev. Lett. {\bf 94} 023202 (2005).

\bibitem{IdziaszekPs} Z. Idziaszek, and T. Calarco,Phys. Rev. Lett. {\bf 96}, 013201 (2006).

\bibitem{Blume} D. Blume, C.H. Greene, Phys. Rev. A {\bf 65}, 043613 (2002).

\bibitem{Bolda} E.L. Bolda, E. Tiesinga, and P.S. Julienne, Phys. Rev. A {\bf 66}, 013403 (2002).

\bibitem{Stoferle} T. St\"oferle {\it et al.}, Phys. Rev. Lett. {\bf 96}, 030401 (2006). 

\bibitem{BoldaQ} E.L. Bolda, E. Tiesinga, and P.S. Julienne, Phys. Rev. A {\bf 68}, 032702 (2003).

\bibitem{Gradshteyn} I.S. Gradshteyn and I.M. Ryzhik, {\it Table of Integrals, Series, and Products} 
(Academic Press, New York, 1965). 

\bibitem{Prudnikov} A.P. Prudnikov, Yu.A. Brychkov, O.I. Marichev, {\it Integrals and Series, vol. II} 
(Gordon and Breach, New York, 1986).

\bibitem{Olshanii} M. Olshanii, Phys. Rev. Lett. {\bf 81}, 938 (1998).

\bibitem{Bergeman} T. Bergeman, M.G. Moore, and M. Olshanii, Phys. Rev. Lett. {\bf 91}, 163201 (2003). 

\bibitem{Block} M. Block, and M. Holthaus, Phys. Rev. A {\bf 65}, 052102 (2002).

\bibitem{Elizalde} E. Elizalde, S.D. Odintsov, A. Romeo, A.A. Bytsenko, and S. Zerbini, {\it Zeta 
Regularization Techniques with Applications} (World Scientific, Singapore 1994). 

\bibitem{Petrov1} D.S. Petrov, M. Holzmann, and G.V. Shlyapnikov, Phys. Rev. Lett. {\bf 84}, 2251 (2000).

\bibitem{Petrov2} D.S. Petrov and G.V. Shlyapnikov, Phys. Rev. A {\bf 64}, 012706 (2000).

\bibitem{Wodkiewicz} A similar regularization operator has been considered in: K. W\'odkiewicz, 
Phys. Rev. A {\bf 43}, 68 (1991).

\bibitem{Moerdijk} A.J. Moerdijk, B.J. Verhaar, and A. Axelsson, Phys. Rev. A {\bf 51}, 4852 (1995).

\bibitem{Timmermans} E. Timmermans, P. Tommasini, M. Hussein, A. Kerman, Phys. Rep. {\bf 315}, 199 (1999).

\bibitem{Goral} K. G\'{o}ral, T. K\"{o}hler, S. Gardiner, E. Tiesinga, and P.S. Julienne, 
J. Phys. B {\bf 37}, 3457 (2004).

\end{thebibliography}
\end{document}